\documentclass[12pt]{article}
\PassOptionsToPackage{breaklinks}{hyperref}
\usepackage{amsmath}
\usepackage{amssymb}
\usepackage{cite}
\newcommand{\ds}{\displaystyle}
\newcommand{\vp}{\varphi}
\newcommand{\tr}{\mbox{tr}(}
\newcommand{\vt}{\vartheta}
\newcommand{\tl}{\tilde{l}}
\newcommand{\dl}{\delta}
\newcommand{\cn}{\text{cn}}
\newcommand{\sn}{\text{sn}}
\newcommand{\dn}{\text{dn}}
\begin{document}
\begin{titlepage}
\begin{flushleft}
DESY 05-219  \\
quant-ph/0510234
 \end{flushleft} \vspace{0.5cm}
\begin{center}
{\Large Quantization of the canonically conjugate  pair  \vspace{0.5cm} \\
 angle and orbital angular momentum}
 \vspace{1.0cm}
\\  {\large H.A.\
Kastrup\footnote{E-mail: Hans.Kastrup@desy.de}} \vspace{0.4cm}
\\ {DESY, Theory Group \\ Notkestr.\ 85, D-22603 Hamburg
\\ Germany} \end{center}
 \begin{center}{\bf Abstract}\end{center}
The question how to quantize a classical system where an angle $\vp$ is one of the basic
canonical variables has been controversial since the early days of quantum mechanics. The problem
is that the angle is a multivalued or discontinuous variable on the corresponding phase space.
The remedy is to replace $\vp$ by the smooth periodic functions $\cos\vp$ and $\sin\vp$. In the
case of the canonical pair $(\vp, p_{\vp})\,,\,p_{\vp}$: orbital angular momentum (OAM), the
 phase space  $\mathcal{S}_{\vp, p_{\vp}}= \{\vp \in \mathbb{R} \bmod{2\pi},p_{\vp}
 \in \mathbb{R}\}$ has the
global topological structure $ S^1 \times \mathbb{R}$ of a cylinder on which the
 Poisson brackets of
the three functions $\cos\vp,\, \sin\vp$ and $p_{\vp}$ obey the Lie algebra of
 the euclidean group
$E(2)$ in the plane. This property provides the basis for the quantization of the system 
in terms of irreducible unitary representations of  the group $E(2)$ or of its
 covering groups. A crucial point is that - due
to the fact that the subgroup $SO(2) \cong S^1$ is multiply connected -  these
representations allow for fractional OAM $l= \hbar(n+\dl),\, n \in \mathbb{Z},\,
 \dl \in [0,1)$. Such
$\dl \neq 0$ have already been observed in cases  like the Aharonov-Bohm  and the fractional
 quantum Hall effects and they correspond to the quasi-momenta of Bloch waves in ideal crystals.
The proposal of the present paper is to look for fractional OAM in
 connection with the quantum optics of 
 Laguerre-Gaussian laser modes in external magnetic fields. The quantum theory of the phase
space  $\mathcal{S}_{\vp,p_{\vp}}$ in terms of unitary representations of $E(2)$
 allows for two types of ``coherent'' states the properties of which are discussed
in detail: Non-holomorphic minimal uncertainty states and holomorphic ones associated
with Bargmann-Segal Hilbert spaces.
\\ \\ PACS number(s): 03.65.Fd, 42.50.-p, 42.55.-f, 78.20.Ls

\end{titlepage}
\section{Introduction and overview}
The problem of quantizing a phase space where an
 angle $\vp \in \mathbb{R} \bmod{2\pi}$ is one
 of the canonical variables has been a controversial issue since the
 founding days of quantum
 mechanics (for a brief historical account see the Introduction of 
 Ref.\ \cite{ka1}). The basic
 reason for the problems is that an angle variable of that type is not a
 smooth periodic
 function on the associated phase space (for details see Appendix A). 

There are two typical (generic) examples where the unit circle $S^1$,
 parametrized by the angle
 $\vp \in \mathbb{R} \bmod{2\pi}$, represents the configuration space, whereas
 the canonically conjugate momentum
 variable $p_{\vp}$ is either
a {\em positive real number} $p_{\vp} >0$\,, i.e.\ $ p_{\vp} \in \mathbb{R}^+ $,
 or a {\em real number}, i.e.\ $p_{\vp} \in \mathbb{R}$.

The ``classical'' representative of a phase space with the 
global topological structure $S^1 \times \mathbb{R}^+$ is the angle - action
 variable description
of the harmonic oscillator: \\ The transformation
 \begin{eqnarray}
 \label{eq:1} q(\vp,I)&=&  \sqrt{2\,I/m\,\omega}\,\cos\vp\,,\,\vp \in 
[0,2\pi)\,,\,I>0\,, \\
 \label{eq:2} p(\vp,I)&=& -\sqrt{2\,m\,\omega\, I}\,\sin\vp\,, \end{eqnarray} is
{\em locally} canonical, i.e.\ it has the property
\begin{equation}dq \wedge dp = d\vp \wedge dI\,, \label{eq:3}  
\end{equation}
and it transforms the Hamilton function 
\begin{equation}
\label{eq:4}  H=\frac{1}{2m}p^2 + \frac{1}{2}m\,\omega^2\,q^2  
\end{equation}
into the simple form $H=\omega \,I$. 

The phase space
\begin{equation}
  \label{eq:5}
  \mathcal{S}_{\vp,I} =\{(\vp,I);\,\vp \in \mathbb{R} \bmod{2\,\pi}\,,\,I \in
 \mathbb{R}^+\}
\end{equation}
 has the {\em global} topological structure of a cone with the tip deleted \cite{ka1}. The 
cone may be parametrized by the 3 functions
\begin{equation}
  \label{eq:6}
  h_0(\vp,I) = I>0 \,,~~ h_1(\vp,I)=I
\,\cos \vp\,,~~ h_2(\vp,I)=-I\,\sin \vp\,,~~~h_0^2-h_1^2-h_2^2=0\,.
\end{equation}
These functions 
obey the Lie algebra $\mathfrak{so}(1,2)$ of the ``proper orthochronous Lorentz group'' 
$SO^{\uparrow}(1,2)$ with
respect to
the Poisson brackets on $\mathcal{S}_{\vp,I}$, namely
\begin{equation}
  \label{eq:7}
  \{f_1,f_2\}_{\vp,I} \equiv \partial_{\vp}f_1\,\partial_If_2 - \partial_If_1\,
\partial_{\vp}f_2\,,~f_j=f_j(\vp,I)\,.
\end{equation}
For the functions \eqref{eq:6} we have 
\begin{equation}
  \label{eq:8}
 \{h_0,h_1\}_{\vp,I}=
-h_2\,,~~\{h_0,h_2\}_{\vp,I}=h_1\,,~~ \{h_1,h_2\}_{\vp,I}=h_0\,.
\end{equation}
This Lie algebra structure on the classical phase space  $\mathcal{S}_{\vp,I}$ serves as the
 basis for the quantization of the system:

  In the quantum theory the corresponding self-adjoint Lie algebra generators $K_0,\,K_1$
 and $K_2$
of certain irreducible unitary representations of the group $SO^{\uparrow}(1,2)$ constitute the
 algebraic basis for the more composite quantum observables of the quantized system like the 
functions $h_0, h_1$ and $h_2$ do for the classical one. What is especially  remarkable is the
 following: 

Obviously it follows from the relations \eqref{eq:1}-\eqref{eq:2} and \eqref{eq:6} that the
 variables $q$ and $p$ may be expressed (non-linearly) by the functions $h_0,h_1$ and $h_2$. 
Similarly: The self-adjoint operators $Q$ and $P$ may be expressed as functions of the
 operators $K_0,K_1$
 and $K_2\,$! Thus, one can replace the basis $\{Q,P,1\}$ of the fundamental
 Weyl-Heisenberg algebra by the basis $\{K_1,K_2,K_0\}$ of the Lie algebra of the
group $SO^{\uparrow}(1,2)$ or one of its (infinitely many) covering groups (for details
 see Ref.\ \cite{ka1}).

Important for the new approach is that the unsuitable variable $\vp \in \mathbb{R} \bmod{2\pi}$
is replaced by the continuous and smooth periodic functions $\cos\vp$ and $\sin\vp$ as
 basic canonical
 variables. This was first suggested in 1963 independently by the physicist
 Louisell \cite{loui} and the mathematician Mackey \cite{mackey}. This makes very good sense,
because any ``decent'' function periodic in $\vp$ may be expanded in a Fourier series where
the terms with $\cos n\vp$ and $\sin n\vp$ can  be expressed as powers of
$\cos\vp$ and $\sin\vp$.

The details of the quantization of the phase space $\mathcal{S}_{\vp,I}$ with its global
 topological structure $S^1 \times \mathbb{R}^+$ in terms of irreducible unitary representations 
of the group $SO^{\uparrow}(1,2)$ and possible associated applications of the somewhat unsual
quantum framework, especially in 
quantum optics, have been discussed elaborately in my 
long paper Ref.\ \cite{ka1}.

The characteristic mechanical example representing a phase space with the global topology
$S^1\times \mathbb{R}$ (a cylinder) is a bead moving frictionlessly on a circular wire with
radius $r_0$ in a horizontal plane.
The position of the bead on the wire is given by an angle $\vp \in \mathbb{R} \bmod{2\pi}$,
 but its angular momentum $p_{\vp}$ may have any value $p_{\vp} \in \mathbb{R}$, positive or
 negative, depending on whether the bead moves anti-clockwise or clockwise. The Hamilton
 function is 
 \begin{equation}
   \label{eq:9}
   H=\frac{p_{\vp}^2}{2\,m\,r_0^2}\,.
 \end{equation}
Whereas the angular momentum $p_{\vp}$ is a constant of motion the value of which depends
 on the 
initial conditions, the angle $\vp$ has the equation of motion
\begin{equation}
  \label{eq:10}
  \dot{\vp} \equiv \omega =\frac{\partial H}{ \partial p_{\vp}} = \frac{ p_{\vp}}{m\,r_0^2} \,,
\end{equation}
with the solution
\begin{equation}
  \label{eq:11}
  \vp(t)= \omega\,t + \vp_0\,.
\end{equation}

The last relation indicates a property of the simple system which plays an important role in our
discussion of its quantum theory below: The position of the bead on the wire can always
 be described by a certain value $\vp \in [0,2\pi)$. But, if not stopped, the bead will pass
that position, e.g.\ $\vp_0$,  many times, namely $n=\omega\,t_0/(2\pi)$ times if it passes 
the ``point'' $\vp_0$ after $t_0>0$ seconds again. Thus, if one looks at the history of the
 motion, the circle is being ``unwrapped'' (arbitrary) many times onto a real line $\mathbb{R}$,
 here represented by the time coordinate $t$. Mathematically speaking, the real line is the
``universal'' covering space of the circle. If the bead circles around twice it runs through a 
2-fold covering, if it circles $q$ times it provides a $q$-fold covering.  

The existence of those covering spaces, especially the universal one, can have  important
consequences for the quantum theory of the system, namely the possibility
of having ``fractional'' or ``quasi orbital angular momenta'' (quasi-OAM),  similar to those
 in the 
Bohm-Aharonov effect, the fractional quantum Hall effect and similar to the quasi-momenta
  associated with Bloch waves in an ideal periodic crystal. It may play
a corresponding important role in the case of the OAM of photons in a
 cylindrical laser beam  (Refs.\ are given below).

The basic functions
\begin{equation}
  \label{eq:15}
 \tilde{h}_1(\vp,p_{\vp}) =  \cos\vp \,,~
\tilde{h}_2(\vp,p_{\vp}) =  \sin\vp\,,~ \tilde{h}_3(\vp,p_{\vp}) = p_{\vp}\,, 
\end{equation}
on the phase space
\begin{equation}
  \label{eq:12}
  \mathcal{S}_{\vp,\,p_{\vp}} = \{s=(\vp,p_{\vp});\,\vp \in 
\mathbb{R}\,\bmod{2\pi},\, p_{\vp} \in \mathbb{R}\,\}
\end{equation}
generate the Lie algebra $  \mathfrak{e}(2)$ of the {\em Euclidean group  $E(2)$ in
the plane\,}:
\begin{equation}
  \label{eq:16}
 \{\tilde{h}_3,\tilde{h}_1\}_{\vp,\,p_{\vp}}=
\tilde{h}_2,~~\{\tilde{h}_3,\tilde{h}_2\}_{\vp,\,p_{\vp}}= -\tilde{h}_1,~~
\{\tilde{h}_1,\tilde{h}_2\}_{\vp,\,p_{\vp}}=0\,.  
\end{equation}
If we characterize the points of a plane by a complex number $z=x+i\,y= r\,e^{i\,\vp}$ then
 the action of the 3-parameter Euclidean transformation group on that plane is given by the
 action of the two subgroups (for more details see Appendix B)
 \begin{eqnarray}
   \label{eq:13}
   \mbox{rotations } R(\alpha):\,\, z &\rightarrow& e^{i\,\alpha}\,z\,,\,\alpha \in
 [0,2\pi)\,\,, \\
\mbox{translations }T_2(t):\,\, z &\rightarrow& z +t\,,\,t = a+i\,b\,,\,a,b \in \mathbb{R}\,.
\label{eq:171} \end{eqnarray}
Like in the case of the phase space \eqref{eq:5} with its Lie algebra structure \eqref{eq:8}
and its quantization in terms of irreducible unitary representations of the group 
$SO^{\uparrow}(1,2)$ or its covering groups, the phase space \eqref{eq:12} can be quantized in
 terms of irreducible unitary representations of the Euclidean group $E(2)$ or its
 covering groups (as to Refs.\ see below): 

In any such irreducible unitary representation the corresponding self-adjoint generators 
$L\,,\,X_1$ and $X_2$ of rotations and translations form the Lie algebra
\begin{equation}
  \label{eq:14}
  \frac{1}{\hbar}\,[L,X_1] = i\,X_2\,,~~~\frac{1}{\hbar}\,[L,X_2] = -i\,X_1\,,~~~[X_1,X_2]
 =0\,.
\end{equation}
The (Casimir) operator
\begin{equation}
  \label{eq:17}
  R^2 = X_1^2+X_2^2
\end{equation}
commutes with all generators of the Lie algebra \eqref{eq:14} and thus has the eigenvalue
$r^2$ in an irreducible unitary representation. Notice that the Lie algebra \eqref{eq:14} 
is invariant 
under the substitution $X_j \rightarrow \gamma \,X_j,\,j=1,2,\, \gamma >0$, so that for
 $R^2 >0$ we can define
the self-adjoint cosine- and sine-operators
\begin{equation}
  \label{eq:18}
  C =\frac{X_1}{\sqrt{R^2}}\,,\,\, S =\frac{X_2}{\sqrt{R^2}}\,,\,\,C^2+S^2 =1\,,
\end{equation}
which obey
\begin{equation}
  \label{eq:19}
 \frac{1}{\hbar}\,[L,C] = i\,S\,,\,~~\frac{1}{\hbar}\,[L,S] = -i\,C\,,~~\,[C,S] =0\,.  
\end{equation}
(Generally the self-adjoint generators of translations are denoted by $P_j$ because they
play the physical role of linear momenta, but here their role is different and that is 
indicated by the notation $X_j$.)

One of the crucial differences between the quantizations of the phase spaces \eqref{eq:5}
and \eqref{eq:12} is that for the former the self-adjoint quantum observable $K_0$, which 
corresponds to the
 positive action variable $I$, has to be a positive definite operator - which it is for the
 positive discrete series of irreducible unitary representations of $SO^{\uparrow}(1,2)$ - ,
whereas in any irreducible unitary representation of $E(2)$ or any of its covering groups
the  generator $L$ of the rotations has arbitrarily large positive {\em and negative}
 eigenvalues!

In order to keep track of the physical dimensions in the following, it is convenient to introduce
the following quantities:
\begin{equation}
  \label{eq:20}
  \mbox{unit of length}:\,\,\lambda_0= \sqrt{\frac{\hbar}{m\,\omega}}\,\,,\,r=\rho\,\lambda_0\,.
\end{equation}
We shall also make frequent use of the dimensionless number
\begin{equation}
  \label{eq:21}
  \epsilon = \rho^{-2} = \frac{\hbar}{m\,\omega\,r^2}\,,
\end{equation}
where in general we shall identify the eigenvalue $r^2$ of the Casimir operator $R^2$ with
the  radius squared $r_0^2$ appearing in Eq.\ \eqref{eq:9}.
The limit $\epsilon \to 0$ characterizes the classical limit $\hbar \to 0$.
(As to physical dimensions: in the applications below the group parameters $a$ and $b$ of Eq.\
 \eqref{eq:171} will have the dimension of an action, just like $p_{\vp}$\,. \\ For $\epsilon=1$
in Eq.\ \eqref{eq:21} we have $\hbar \omega = O(\text{eV})$ if $m \approx m_e$
 and $r=10^{-10}\text{m}$.)

The irreducible unitary representations of the Euclidean group $E(2)$ and its covering groups
may all be implemented in a Hilbert space $L^2(S^1,d\vp/2\pi)$ of functions $\psi(\vp)$
 with the scalar product
\begin{equation}
  \label{eq:22}
  (\psi_2,\psi_1) = \int_0^{2\pi}\frac{d\vp}{2\pi}\,\psi^*_2(\vp)\psi_1(\vp)\,.
\end{equation}
The  irreducible unitary representations are in general characterized by two real numbers,
namely by the pair \cite{ish} (see also the literature quoted in
Appendix B)
\begin{equation}
  \label{eq:23}
  (\rho,\delta)\,,\, \rho >0\,,\,\delta \in [0,1)\,.
\end{equation}
The representations themselves are given by 
\begin{eqnarray}
  \label{eq:24}
  [U^{\rho,\delta}(\alpha)\psi](\vp)&=& e^{\ds -i\,\delta\,\alpha}\psi[(\vp-\alpha)
 \bmod{2\pi}]\,,\\
\label{eq:25} [U^{\rho,\delta}(t=a+ib)\psi](\vp)&=&e^{\ds -(i/\hbar)\rho\,(a\,\cos\vp+b\,\sin\vp)}
\psi(\vp)\,.
\end{eqnarray}
The parameter $\delta$ differentiates between the irreducible unitary representations of the
different covering groups: 

For the irreducible unitary representations of the group $E(2)$ itself we have $\delta =0$.
 If $\delta$ equals a rational number $p/q$,  with
$p,q \in \mathbb{N}$ and no common divisor, then we have a representation of a q-fold
 covering of $E(2)$ (see below) and if $\delta$ is an irrational number we have a 
 representation of
the universal covering group $\tilde{E}(2)$.

If we define the self-adjoint generators of the 1-dimensional subgroups corresponding to
 the parameters
$\alpha$ and $t = a+i\,b$ by
\begin{equation}
  \label{eq:26}
  U^{(\rho,\delta)}(\alpha)=e^{\ds -(i/\hbar)L_{\delta}\,\alpha}\,\,,\,\,
U^{(\rho,\delta)}(t)=e^{\ds-[i/(\hbar\lambda_0)]\,(X_1a+X_2b)}\,,
\end{equation}
we obtain
\begin{equation}
  \label{eq:27}
  \frac{1}{\hbar}L_{\delta} \equiv \tilde{L}_{\delta}= \frac{1}{i}\partial_{\vp} +
 \delta\,\,,\,\,X_1 = r\,\cos\vp\,\,,\,\,X_2 = r\,\sin\vp\,.
\end{equation}

The Hilbert space $L^2(S^1,d\vp/2\pi)$ with the scalar product \eqref{eq:22} has the orthonormal 
basis
\begin{equation}
  \label{eq:28}
e_n(\vp)=e^{\ds i\,n\,\vp}\,\,,\,n \in \mathbb{Z}\,.  
\end{equation}
The functions \eqref{eq:28} are eigenfunctions of the OAM-operator $L_{\delta}$:
 \begin{equation}
   \label{eq:29}
   L_{\delta}\,e_n = \hbar\,(n+\delta)\,e_n\,,\,n \in \mathbb{Z}\,.
 \end{equation}
What appears surprising is the fact that the OAM-operator can have non-integer eigenvalues.
This is typical for the rotation group $SO(2)$ (or $U(1)$) which has the non-trivial topological
structure of the circle $S^1$ and the additive group $\mathbb{R}$ of the real numbers as its 
universal covering group $\widetilde{SO(2)}$. 

With the integers $\mathbb{Z}$ as an abelian subgroup of $\mathbb{R}$ 
we may write 
\begin{equation}
  \label{eq:30}
  SO(2) \cong S^1 \cong \widetilde{SO(2)}/(2\pi\mathbb{Z}) = \mathbb{R}/(2\pi\mathbb{Z})\,,
\end{equation}
which is just another way of writing $\vp \in \mathbb{R} \bmod{2\pi}$. 

As the group $\mathbb{Z}$ is abelian, all its irreducible unitary representations are
 1-dimensional:
 \begin{equation}
   \label{eq:31}
   \mathbb{Z} \to \{e^{\ds -i 2\pi n \delta}\,,\,n \in \mathbb{Z}\,\}\,,\,\delta \in [0,1)\,.
 \end{equation}
The numbers $\delta$ characterize the different representations of $\mathbb{Z}$.
This is the deeper reason for the appearence of the additional parameter $\delta$ 
in the transformation formula \eqref{eq:24} and in the eigenvalue equation \eqref{eq:29}.
The mathematical background is very thoroughly discussed in Ref.\ \cite{ish}.

A further essential mathematical remark is: the first homotopy group $\pi_1$ of $S^1$ 
coincides with $\mathbb{Z}$, too! This shows the non-trivial topological structure of the
circle $S^1$ which is not simply path-connected!

Different $\delta$ lead to different spectra of $L_{\delta}$ and therefore such operators
 are not unitarily equivalent.

In the discussion above we have assumed that the different irreducible unitary representations 
corresponding to different $\delta$ are all realized in the same Hilbert space with the basis
\eqref{eq:28}. By making the unitary transformations
\begin{equation}
  \label{eq:40}
  e_n(\vp)=e^{\ds in\vp} \to e_{n,\delta}(\vp)=e^{\ds i\delta\vp}e_n(\vp)=e^{\ds i(n+\delta)\vp}
\,\,\,
\forall\, n \in \mathbb{Z}\,,
\end{equation}
we can define a separate Hilbert space $L^2(S^1,d\vp/2\pi,\delta)$ for each $\delta$. In these
Hilbert spaces the generators \eqref{eq:27} now have the common form
\begin{equation}
  \label{eq:41}
  \frac{1}{\hbar}L_{\delta} \equiv \tilde{L}_{\delta}= \frac{1}{i}\partial_{\vp} 
 \,\,,\,\,X_1 = r\,\cos\vp\,\,,\,\,X_2 = r\,\sin\vp\,,
\end{equation}
i.e. now the operators are independent of $\delta$, the dependence of which is shifted to the
basis \eqref{eq:40}. The basis functions \eqref{eq:40} - and any function $\psi(\vp)$ expanded
 with
respect to them - obey the boundary condition 
\begin{equation}
  \label{eq:42}
  e_{n,\delta}(\vp + 2\pi) = e^{\ds i2\pi\delta}\,e_{n,\delta}(\vp)\,\,,\,\psi(\vp + 2\pi) =
e^{\ds i2\pi\delta}\psi(\vp)\,\,,\,\psi \in L^2(S^1,d\vp/2\pi,\delta)\,.
\end{equation}
The spectrum \eqref{eq:29} of $L_{\delta}$  does not change,
 of course, due to the unitarity of the transformation! An additional mathematical interpretation
of the phase angle $\dl$ in terms of self-adjoint extensions of a symmetric operator is briefly
discussed in Appendix A.

Consider now the case $\dl = p/q\,,\,p,q \in \mathbb{N}$ and divisor-free, mentioned above. Then
\begin{equation}
  \label{eq:297}
  \psi(\vp+q\,2\pi)=\psi(\vp)\,.
\end{equation}
This property characterizes the unitary representation of a $q$-fold covering of $SO(2)$.

A value $\delta \neq 0$ can have significant physical consequences: 

i) The Hamilton operator (discussed in Refs.\ \cite{schul1,asor1})
\begin{equation}
  \label{eq:32}
  H_{\delta}=\frac{L_{\delta}^2}{2mr^2} = \frac{1}{2}\,\epsilon\, \hbar \omega
 (\frac{1}{i}\partial_{\vp}+ \delta)^2
\end{equation}
has the eigenfunctions \eqref{eq:28} and the eigenvalues
\begin{equation}
  \label{eq:33}
  E_{n,\delta} = \frac{1}{2}\,\epsilon \, \hbar \omega (n+\delta)^2\,\,,\, n \in \mathbb{Z}\,.
\end{equation}
The ground state energy is given by
\begin{equation}
  \label{eq:34}
  E_{n=0,\delta} = \frac{1}{2}\,\epsilon\, \hbar \omega \,\delta^2~~~ \mbox{or}~~
 E_{n=-1,\delta} = \frac{1}{2}\,
\epsilon\, \hbar \omega \,(1-\delta)^2\,,
\end{equation}
depending on whether $\delta \in [0,1/2)$ or $\delta \in (1/2,1)$\,. For $\delta = 1/2$ the
 ground state is degenerate.

Alternatively one may discribe the system in terms of the operators \eqref{eq:41} and
the eigenfunctions \eqref{eq:40}. All the physical consequences are the same, because of the
unitary equivalence of the two descriptions.

ii) The expression 
\begin{equation}
  \label{eq:35}
  L_{\delta} = \hbar\,(\frac{1}{i}\partial_{\vp} + \delta)
\end{equation}
equals the ``covariant derivative'' 
\begin{equation}
  \label{eq:36}
  \frac{\hbar}{i}\partial_{\vp} + q\,\hat{A}_{\vp}\,\,,\,\hat{A}_{\vp} = \frac{\Phi}{2\pi}\,,
\end{equation}
associated with the  component  $A_{\vp}=\hat{A}_{\vp}/r$
 of the vector potential (in cylindrical coordinates) surrounding the thin line of a
 magnetic flux $\Phi$ which causes the
Aharonov-Bohm effect for  particles of charge $q$ \cite{ahar1}. \\ In this case we have
\begin{equation}
  \label{eq:37}
  \delta = \frac{q}{2\pi\,\hbar}\Phi\,
\end{equation}
and the Hamiltonian \eqref{eq:32} takes the form
\begin{equation}
  \label{eq:39}
  H= \frac{1}{2}\,\epsilon\, \hbar \omega (\frac{1}{i}\partial_{\vp}+ \frac{q}{2\pi\hbar}\Phi)^2.
\end{equation}
For single  electrons we have $q=-e_0$. 

The observable phase shift $\Delta \theta$ responsible for the change in the interference pattern
caused by the presence of the magnetic flux $\Phi$ is
\begin{equation}
  \label{eq:38}
  \Delta \theta = 2\pi\delta = \frac{q}{\hbar}\,\Phi\,.
\end{equation}
The change of the interference pattern ceases for $\delta =1$, which defines a flux quantum
$\Phi_0(q) = h/q$ (or $h/|q|$) associated with a charge $q$. $q=-2e_0$ yields the fundamental
 flux quantum
 $\Phi_0= h/(2e_0)$ of superconductivity. 

The interpretation of the Aharonov-Bohm effect in terms of unitary representations of the 
universal covering group of $SO(2)$ (or $E(2)$) has been discussed - at least in principle -
by a number of authors
\cite{schul2,mart,breit,moran1,moran2,lands1}, most explicitly first by C.\ Martin \cite{mart}.

iii)  The relations \eqref{eq:36} and \eqref{eq:38} reflecting the 
 Aharonov-Bohm effect was the stimulating example
for the flourishing of the concept ``anyons'' and their
``fractional'' statistics \cite{lei,wil1} from 1982 on (see the reviews
 \cite{gif,lus,wil2,jain,fort,lerd,khare,laugh}): 

 If one considers the particle with charge $q$ (spin $s$) and the magnetic flux $\Phi$ which
 influences it 
as a new (fictitious) composite entity with ``charge'' $\theta =2\pi \delta = q\Phi$ confined
 to a plane perpendicular to the straight flux line, then this
 object can be viewed as having an angular momentum $\hbar(n+\delta)$. Assume that one has two
 identical such
objects localized at different positions. If $\vp$ is the polar angle of the vector $\vec{x}_2
 -\vec{x}_1$   connecting the two objects, the wave function $\psi(1,2;\vp)$ of the relative
 motion
should be symmetric if the two objects are bosons ($\delta =0$) and antisymmetric for fermions
($\delta = 1/2$). Interchange of the two is implemented by the substitution $\vp \to \vp + \pi$. 
So the correct behaviour of the wave functions for bosons and fermions is guaranteed by the
property
 \begin{equation}
   \label{eq:45}
   \psi(2,1;\vp+\pi) = e^{\ds i2\pi\delta}\psi(1,2;\vp)\,\,.
 \end{equation}
For $\delta \neq 0,\,1/2$ the last equation defines a new kind of statistics and the associated
objects are called ``anyons'' and the corresponding statistics ``fractional''. As to the
 associated braid group symmetry (replacing the usual permutation group) see the reviews quoted
above. There is strong evidence that anyons play a crucial role in the description of the
Fractional Quantum Hall Effect (see reviews). There the parameter $\dl=p/q$ represents the
filling factor $\nu$ of the Landau levels which becomes fractional due to certain collective
mechanisms.

iv) There is still another very interesting physical interpretation of the real numbers
 $\delta$: \\
The situation here is completely analogous to that for Bloch wave functions of an ideal crystal
\cite{kitt,zim}:

 Assume an infinitely long 1-dimensional ideal crystal with lattice
 constant
 $a >0$. Then it follows from group theory that the wave function $\psi_k(x)$ of a particle
 moving
in such a lattice under the influence of a periodic potential $V(x+a)=V(x)$ has the general
form
\begin{equation}
  \label{eq:43}
  \psi_k(x)= e^{\ds ikx}u_k(x)\,\,,\,u_k(x+a)=u_k(x)\,.
\end{equation}
The ``reduced'' or ``crystal'' wave vector $k$ lies in the interval $[0,2\pi/a)$ (or $[-\pi/a,
\pi/a)$), called the ``first Brillouin zone'' of the reciprocal lattice. The quantity $\hbar
 k$ is usually called the 
``quasi-momentum'' of the particle. The wave function \eqref{eq:43} has the property
\begin{equation}
  \label{eq:44}
  \psi_k(x+a) = e^{\ds ika}\psi_k(x)\,,
\end{equation}
which is to be compared with the relations \eqref{eq:42}. The periodicity in $a$ of the function
$u_k(x)$ in Eq.\ \eqref{eq:43} corresponds to the periodicity in $2\pi$ of the $e_n(\vp)$!
If we ignore for a moment that $x$ has the dimension of a length and put $a=2\pi$, we can
`` identify'' $\delta$ with $k$ and the interval $[0,1)$ from Eq.\ \eqref{eq:23} becomes
the ``first Brillouin'' zone of our orbital angular momentum problem! \\
It is obvious from the long experiences in solid state physics that a boundary condition \\
$\psi(x+a) = \psi(x)$ would be completely inappropriate, because it would mean $k=0$\,!
Therefore one should be similarly careful with the use of the ``quasi-OAM''
$\mu=\hbar\,\delta$! The analogies between  fractional OAM and  quasi-momenta of Bloch waves
were first pointed out by Schulman \cite{schul3,schul2} (as to Zak's related work see Ref.\
\cite{zak}  below).

v) Finally there is the related so-called ``$\theta$-vacuum'' structure of QCD associated with
an additional $U(1)$-symmetry and its (fractional) representations
 (see the reviews \cite{trei,bala,weinb}).
The relationship between this feature and Bloch waves has been pointed out by Jackiw 
\cite{jack1}. 

 It is to be stressed that   non-trivial quasi-OAM $\delta$ can only be
attributed to  orbital angular momenta associated with covering groups of the group $SO(2)$,
i.e.\ to systems with {\em cylindical} symmetries! They appear also in systems with symmetries
where the group $SO(2)$ is a maximal compact subgroup, e.g. for $SO^{\uparrow}(1,2)$ or its
inhomogeneous generalization $ISO^{\uparrow}(1,2)$ (Poincar\'{e}  group in one time and
 two space dimensions). \\
 They are not possible
for the spatial rotation group $SO(3)$ the universal covering group of which is the double 
covering $SU(2)$ that allows only for integer and half-integer angular momenta!

The following remark is  important: 

{\em  Fractional OAM $\dl$ violate $T$- and $P$-invariance}, except for $\dl=0$ (bosons) and
$\dl=1/2$ (fermions) in the following sense: Classically we may write $p_{\vp} = x\,p_y-
y\,p_x$\,. This implies that we have the transformations
\begin{eqnarray}
  \label{eq:298}
 \text{time reversal}~T\,&:&\, p_{\vp}\to -p_{\vp}\,,\\
\text{space reflection}~P\,(x\to -x,\,y\to y)&:&\, p_{\vp} \to -p_{\vp}\,. \label{eq:299}
\end{eqnarray}
The quantum mechanical version of the transformation $T$  here is simply implimented by
 complex conjugation of the
 wave function:
\begin{equation}
  \label{eq:300}
  T:~e_{n,\dl}(\vp)=e^{\ds i(n+\dl)\vp} \to e_{n,\dl}^*(\vp)=e^{\ds-i(n+\dl)\vp}\,.
\end{equation}
For the space reflection $P$ we get
\begin{equation}
  \label{eq:301}
  P:~(n+\dl) \to -(n+\dl)\,,
\end{equation}
so that the product $P\circ T$ leaves the wave function $e_{n,\dl}(\vp)$ invariant.
But the transformations \eqref{eq:300} and \eqref{eq:301} cannot be implemented separately
 within a given representation \eqref{eq:24}, because now we have instead of Eqs.\ \eqref{eq:42}:
\begin{equation}
  \label{eq:302}
  e_{n,\dl}^*(\vp+2\pi)= e^{\ds-i2\pi\dl}\, e_{n,\dl}^*(\vp)\,,~~\psi^*(\vp+2\pi)=e^{\ds -i2\pi
\dl}\,\psi^*(\vp)\,.
\end{equation}
These complex conjugate functions may be associated with a representation
 $U^{(\rho,\dl')}(\alpha)$, where
 \begin{equation}
   \label{eq:303}
   T\,\, \text{or}\,\,P:~\dl \to \dl'=1-\dl\,.
 \end{equation}
So only for $\dl =0,\,1/2$ we can implement $T$ and $P$ within the same irreducible unitary
representation. Otherwise these symmetries are violated! 

This property suggest that fractional OAM should be possible in $T$- or $P$-violating systems.
Such systems are realized if an {\em external magnetic field} $\vec{B}$ is applied which 
violates $T$-invariance ($\vec{B}$ changes sign under $T$). Examples can be seen above:
 The Bohm-Aharonov effect \eqref{eq:37}
and the fractional quantum Hall effect are both associated with external magnetic fields!

Another $T$-invariance breaking experimental possibility is to impose an appropriate external OAM
$L^{ext}_{\vp}$ by rotating the system.

I now come to an important point of the present paper: The possibility of fractional OAM
in quantum optics:

Since 1992 \cite{woerd1} there is an increasing number of papers dealing with theory and 
experiments (see the reviews \cite{allen1,jopt,barn1,sant} and some more recent papers
 \cite{barn2})
 concerning {\em orbital} angular momenta (OAM) of photons in so-called
 ``Laguerre-Gaussian'' laser modes, cylinder symmetrical laser beams the (classical)
 amplitudes of which
contain the azimuthal angle- and OAM-dependent factors
\begin{equation}
  \label{eq:46}
 r^{\ds|\tilde{l}|}\, e^{\ds i \tilde{l}\vp}\,,\,\tilde{l} \in \mathbb{Z}\,,
\end{equation}
where $r$ is the radial variable of the cylinder coordinates.

Whereas the spin of the photon provides only a 2-dimensional state space for the study of 
quantum information problems, questions of entaglement etc.\ etc., 
its OAM  provides - at least in principle - one which can have an arbitrarily
high dimension. If implementable this would lead to a wealth of new theoretical, experimental
and even technological possibilities. 

The prominent question in the present context  is, however, whether one can find
 fractional OAM of the
photon and separating their properties from those of the photon spin (a problem still under
discussion). In  order to obtain such quasi-OAM one probably needs a T-violating environment
such as an external magnetic field or an external OAM. So one has to look for
 an ``OAM Faraday effect'' \cite{land,somm,barron,handb,bud}, similar magneto-optical
 phenomena or for mechanical-optical effects. 

One sees from Eq.\ \eqref{eq:46} how the laser beam amplitudes get modified in the neighbourhood
of the central beam axis if the integer $\tilde{l}$ is replaced by a non-integer $\tilde{l} +
\delta $. Pictures of the  intensity distributions of certain Laguerre-Gaussian laser
 modes in planes transversal to the central
 axis of the beam can be found in Refs.\ \cite{barn2}.

Experimental generation of fractional OAM has recently been discussed in Ref.\ \cite{oem}. 

One question is whether one should expect a continuous $\dl$, a rational or a discontinuous one
 in such optical
experiment. Again, the Bohm-Aharonov, the fractional quantum Hall effect and the Bloch waves
 respectively,
 indicate the direction into which to look: the Bohm-Aharonov effect has a continuous $\dl$, 
essentially given by the value of the external flux $\Phi$. The system consists of moving free
charged particles with no collective interactions. In the case of the fractional quantum Hall
effect with its rational $\dl=p/q$ the collective dynamics for the electron creates some
 sort of quantum fluid  and therefore the situation is qualitatively different. In the case of
Bloch waves the quasi-momentum $\hbar k$ of a ``free'' electron can be affected by complicated
interactions with the ions of the lattice, leading to energy gaps etc., but also by the influence
 of external magnetic fields \cite{kitt,zim}.  

The first two examples suggest that   one probably should expect a continuous $\dl$
in the quantum optical setups used up to now (see Refs.\ given above) {\em plus} an
 external magnetic field.
 But at very low temperatures, with an appropriate
medium causing the magneto-optical ``fractionizing'', the situation may be different and turn
``rational''!

In order to have a proper theoretical description of all quantum aspects involved one
needs a satisfactory quantization of the phase space \eqref{eq:12}. It is the purpose of the
 present paper to draw attention - especially that of the quantum optics community - to the 
existing group theoretical quantization of the phase space \eqref{eq:12} in terms of the
Euclidean group $E(2)$ and (or) its covering groups, the associated coherent states
and their uncertainty relations
 etc. I shall draw on previous work by other authors, but
 is my  emphasis is on the {\em  physical 
possibility of fractional orbital angular momenta}.

Sec.\ 2 explains und summarizes the group theoretical quantization of the phase space 
\eqref{eq:12}.

 Sec.\ 3 discusses a class of coherent states derivable from a minimal
uncertainty requirement. These coherent states form a complete set, but are not holomorphic
in the pair $(\vp,p_{\vp}) $. 

Coherent states holomorphic in $(\vp,p_{\vp})$ may be generated
 in two ways:

Sec.\ 4: Applying a mapping, introduced in mathematics by Weil \cite{weil}
 and in physics independently
by Zak \cite{zak} in connection with Bloch waves, one can ``periodize'' the real
 part $q/\lambda_0 $ of
 the complex number  $z= q/\lambda_0+i\lambda_0\,p/\hbar$ ($\lambda_0$: see Eq.\ \eqref{eq:20})
occurring in the usual Schr\"{o}dinger-Glauber coherent states. This procedure leads 
automatically to the introduction of the fraction $\delta$. Thus, starting from well-known
coherent states one can construct corresponding ones for the group $E(2)$ and its covering
 groups. \\
This may even lead to a possible experimental generation of the new coherent states, if one
could construct such ``periodizers'' for the standard coherent states experimentally!

Sec.\ 4.2 discusses expectation values and fluctuations of the basic observables $C,\,S$ and
$L$ with respect to these holomorphic coherent states.

Sec\, 5: The same coherent states may be generated by a certain complexifiction of the group
$SO(2)$ introduced for compact groups by Hall \cite{hall1} (more Refs.\ will be given below).
 In this 
approach the coherent states can be generated as eigenstates of a certain annilation operator
$B$ (a nonlinear function of the generators of $E(2)$) with complex eigenvalues
 $e^{\ds -iz}\,,z=\theta +i\,\tilde{l}\,,\,\theta \in \mathbb{R} \bmod{2\pi},\, 
\tilde{l} \in \mathbb{R}$\,.

 The commutation relation $B^{\dagger}B=q\,BB^{\dagger},\,q=e^{-2\,\epsilon}$,
may be rewritten as $aa^{\dagger}-q\,a^{\dagger}a = q^{-N}$. That is, the operators $B$
 and $B^{\dagger}$ generate
a $q$-deformed Born-Dirac-Heisenberg-Jordan algebra  which, perhaps, may be tested in quantum
optics, too.  

Sec.\ 6 discusses the time evolution of both types of coherent states with respect to the
Hamiltonian \eqref{eq:32}.

Appendix A summarizes the problems associated with a conventional quantization of the
canonical variable ``angle'', Appendix B sketches its quantization (together with the canonically
conjugate $p_{\vp}$) in terms of unitary representations of the group $E(2)$. Appendix C
 contains some properties of Jacobi's $\vartheta$-functions needed in the main text.

\section{The Euclidean group $E(2)$ as the canonical group of the phase space 
$ \mathcal{S}_{\vp,p_{\vp}}$}
That the phase space \eqref{eq:12} has something to do with the Euclidean group $E(2)$
 can already be
seen from the Lie algebra \eqref{eq:16} of that group generated by the basic observables
\eqref{eq:15}. For the Euclidean group $E(2)$ to be the so-called ``canonical group'' of
 the phase
space \eqref{eq:12} it should fulfill a number of properties (see Refs.\ \cite{ish,stern},
  Appendix A of Ref.\ \cite{ka1} and Appendix B of the present paper): 

 The group action 
\begin{equation}
  \label{eq:47}
 s \equiv (\vp,p_{\vp}) \to s' \equiv (\vp',p_{\vp}') = g_{\alpha,t}[(\vp,p_{\vp})]\,,~t=a+ib\,,
\end{equation}
i) should be symplectic,
\begin{equation}
  \label{eq:48}
  d\vp'\wedge dp_{\vp}' = d\vp \wedge dp_{\vp}\,.
\end{equation}
 ii) It should be ``transitive'', i.e.\ given any two points $s_i \in 
\mathcal{S}_{\vp,p_{\vp}}\,,\,i=1,2$, then there exists a transformation $g_{\alpha,t}[...]$ 
which
maps one point onto the other. \\
iii) It should be effective [or almost effective], i.e.\ if a transformation
 $g_{\alpha,t}[...]$ leaves
all points $ s \in \mathcal{S}_{\vp,p_{\vp}}$ invariant, then $g_{...}$ is the identity
element $(\alpha=0,a=0,b=0),$ [or $g_{...}$ is an element of a discrete abelian subgroup
 $\subset \mathbb{Z}$ of the center $\mathbb{Z}$ of the universal covering group
 $\tilde{E}(2)$]. \\
iv) The 1-parameter transformation subgroups induced by  group elements
\begin{equation}
  \label{eq:49}
  g_{\gamma} = e^{\ds -A\,\gamma}\,,\,\gamma \in \mathbb{R}\,,\, A \in \mbox{ Lie algebra
 $ \mathfrak{e}(2)$}\,, \end{equation} generate vector fields $\breve{A}(s)$ on
 $\mathcal{S}_{\vp,p_{\vp}}$: If $f(s)$ is a smooth function, then $g_{\gamma}$ generates
 \begin{equation}
   \label{eq:60}
   [\breve{A}f](s) = \lim_{\gamma \to 0} \frac{1}{\gamma}[f(e^{\ds -A\gamma}s)-f(s)]\,.
 \end{equation}
Such vector fields generally have the form
\begin{equation}
  \label{eq:61}
  a_{\vp}(\vp,p_{\vp})\,\partial_{\vp}+a_{p_{\vp}}(\vp,p_{\vp})\,\partial_{p_{\vp}}\,.
\end{equation}
But as the transformations \eqref{eq:49} are symplectic, the vector fields \eqref{eq:60}
induced by them are locally Hamiltonian, i.e.\ there exists a function $f(\vp,p_{\vp})$ such
 that locally
\begin{equation}
  \label{eq:62}
  a_{\vp}= -\partial_{p_{\vp}}f\,,\,\, a_{p_{\vp}} = \partial_{\vp}f\,.
\end{equation}
The three vector fields $\breve{X}_1\,,\,\breve{X}_2$ and $\breve{L}$ induced by the three
1-parameter subgroups of $E(2)$ associated with the parameters $a,\,b$ and $\alpha$ 
obey the Lie algebra $\mathfrak{e}(2)$,
\begin{equation}
  \label{eq:59}
  [\breve{L},\breve{X}_1] = \breve{X}_2\,,\,\,[\breve{L},\breve{X}_2] =- \breve{X}_1\,, \,\,
[\breve{X}_1, \breve{X}_2] = 0\,.
\end{equation}
v) Crucial is finally that the three Hamiltonian functions $f_i\,,\,i=1,2,3,$
 corresponding to the three induced vector fields $
   \breve{X}_1\,,\,\breve{X}_2$ and $\breve{L}$ 
  are {\em globally} defined on $\mathcal{S}_{\vp,p_{\vp}}$ and obey the Poisson
 bracket Lie algebra  $\mathfrak{e}(2)$:
  \begin{equation}
    \label{eq:54}
   \{f_3,f_1\}_{\vp,\,p_{\vp}}=
f_2,~~\{f_3,f_2\}_{\vp,\,p_{\vp}}= -f_1,~~
\{f_1,f_2\}_{\vp,\,p_{\vp}}=0\,.   
  \end{equation}

All the above required  properties i) - v) are fulfilled by the following transformation
 law (see Ref.\ \cite{ish2} and Appendix B):
\begin{equation}
  \label{eq:55}
  g_{\alpha,t}[(\vp,p_{\vp})] =(\vp',p_{\vp}')=[(\vp+\alpha)\bmod{2\pi},\, p_{\vp}+ a\, \sin
(\vp+\alpha) -b\,\cos(\vp +\alpha)]\,.
\end{equation}
According to Eq.\ \eqref{eq:60} the vector fields $\breve{X}_1,\,\breve{X}_2$ and
 $\breve{L}$ can be read off  the Taylor expansion of
 \begin{equation}
   \label{eq:56}
   f[\vp-\alpha,\,p_{\vp}-a\,\sin(\vp-\alpha)+b\,\cos(\vp-\alpha)]
 \end{equation}
with respect to $a,\,b$ and $\alpha$:
\begin{equation}
  \label{eq:57}
  \breve{X}_1 = -\sin \vp\, \partial_{p_{\vp}}\,,~~ \breve{X}_2 = \cos \vp\,
 \partial_{p_{\vp}}\,,~~
\breve{L} = -\partial_{\vp}\,.
\end{equation}
The associated global Hamiltonian functions according to \eqref{eq:62} are
\begin{equation}
  \label{eq:58}
  f_1(\vp,p_{\vp}) = \cos \vp\,,\,\,f_2(\vp,p_{\vp}) = \sin \vp\,,\,\, f_3(\vp,p_{\vp})= p_{\vp}
\,, 
\end{equation}
which are just the basic classical observables \eqref{eq:15} we started from!

All these group theoretical features as to the classical phase space \eqref{eq:12} form the
basis for its consistent quantization, completely similar to those of the
 ``Born-Dirac-Heisenberg-Jordan-Weyl'' group of the usual phase space $\mathcal{S}_{q,p} =
 \{(q,p) \in
\mathbb{R}^2 \}$ (see, e.g.\ the Refs.\ \cite{ish}\cite{ka1}):

The main elements of the quantization scheme have already been discussed in the introduction: 

In the quantum theory the classical basic observables \eqref{eq:15} with their $E(2)$ Lie algebra
structure \eqref{eq:16} become the self-adjoint generators \eqref{eq:19}, or - explicitly - 
\eqref{eq:27} with $r=1$, in an irreducible unitary representation. At first sight the 
quantization does not appear to be unique: according to Eqs.\ \eqref{eq:24} and \eqref{eq:25}
each irreducible unitary representation depends on two parameters $\rho$ and $\delta$. 
 
However,
the parameter $\rho$ represents the freedom of having different numerical values for Planck's
constant - see Eq.\ \eqref{eq:21} - depending on the system of units  employed!
One has the same type of freedom in the conventional quantization scheme with its Weyl-
Heisenberg group and the associated von Neumann - Stone uniqueness theorem \cite{ish}.

 The parameter $\delta$ is new, however. In the introduction we have seen that it is
a quantum manifestation of the fact that the group $SO(2)$ (or $U(1)$) has an infinite number
of different covering groups each of which can be characterized by its non-trivial center $Z_q$
generated by $e^{2\pi\,i/q}$. Or, in other words, the {\em appearance of the parameter $\delta$
is a quantum effect of the non-trivial topology of the unit circle} parametrized by the
angle $ \phi \in [0,2\pi)$! I have emphasized in the introduction that there are a number of
important physical examples which show consequences of such a non-trivial topology. Therefore,
there might be more consequences of that topology in physics than we are aware of up to now.

From the Eqs.\ \eqref{eq:27} we obtain the explicit form
of the self-adjoint operators $C,\,S$ and $L_{\dl}$:
\begin{equation}
  \label{eq:50}
 \frac{1}{\hbar}L_{\delta} \equiv \tilde{L}_{\delta}= \frac{1}{i}\partial_{\vp} +
 \delta\,\,,\,\,C = \cos\vp\,\,,\,\,S = \sin\vp\,,  
\end{equation}
in the Hilbert space with the scalar product \eqref{eq:22} and the basis \eqref{eq:28}.

Equivalently one may use the operators
\begin{equation}
  \label{eq:51}
\frac{1}{\hbar}L_{\delta} \equiv \tilde{L}_{\delta}= \frac{1}{i}\partial_{\vp} 
 \,\,,\,\,C = \cos\vp\,\,,\,\,S = \sin\vp\,,   
\end{equation}
in a Hilbert space with the basis \eqref{eq:40} for functions with the boundary condition
\eqref{eq:42}.

The operators \eqref{eq:50} - or \eqref{eq:51} - obey the commutation relations
\begin{equation}
  \label{eq:52}
   \frac{1}{\hbar}\,[L_{\delta},C] = i\,S\,,~~\frac{1}{\hbar}\,[L_{\delta},S] =
 -i\,C\,,~~[C,S] =0\,.
\end{equation}
The last commutator shows that $\cos \vp$ and $\sin \vp$ may be measured simultaneously,
leading to a unique value $\vp \in [0,2\pi)$\,! This is not so in the quantum theory
of the phase space \eqref{eq:5}, where the self-adjoint operators $K_1$ and $K_2$ corresponding
to the basic functions $h_1$ and $h_2$ from Eqs.\ \eqref{eq:6} with their Lie algebra structure
\eqref{eq:8} do not commute!

In the following discussions it is convenient to work with the ``dimensionless'' operator
$\tilde{L}$ instead with $L$ itself. It is always possible to restore the associated 
$\hbar$-dependence in the formulae.

  If $A,\,B$ are any two of the three self-adjoint operators \eqref{eq:50} or
 \eqref{eq:51} and 
$\psi$ an element of their domain of definition, then we have the general uncertainty relation
\cite{rob,schro,merz,jack2},\cite{ka1} 
\begin{equation}
  \label{eq:63}
  (\Delta A)^2_{\psi}\,(\Delta B)^2_{\psi} \geq  |\langle S_{\psi}(A,B)\rangle_{\psi}|^2 
+\frac{1}{4}|\langle [A,B]\rangle_{\psi}|^2\,,
\end{equation}
for the mean square deviations
\begin{equation}
  \label{eq:53}
(\Delta A)^2_{\psi} = \langle (A-\langle A \rangle_{\psi})^2\rangle_{\psi}\,,\,\, \mbox{ with }
 \langle A
\rangle_{\psi}
 \equiv (\psi,A\psi)\,,   
\end{equation}
where
\begin{equation}
  \label{eq:64}
  S_{\psi}(A,B) = \frac{1}{2}(AB+BA)-\langle A\rangle_{\psi}\,\langle B\rangle_{\psi}\,.
\end{equation}
Of special interest for applications are those states $\psi_0$ for which the relation
\eqref{eq:63} becomes an equality (so-called ``minimal uncertainty states''). Equality holds
iff
\begin{equation}
  \label{eq:65}
  (B-\langle B \rangle_{\psi_0})\psi_0 = \sigma\, (A-\langle A \rangle_{\psi_0})\psi_0\,\,,\,
~~\sigma= \gamma -i\,s \in \mathbb{C}\,.
\end{equation}
The real numbers $\gamma$ and $s$ are given by
\begin{equation}
  \label{eq:66}
  \gamma =\frac{\langle S_{\psi_0}(A,B) \rangle_{\psi_0}}{(\Delta A)^2_{\psi_0}}\,,\,\,
~~ s= \frac{i}{2}\, \frac{
\langle [A,B] \rangle_{\psi_0}}{(\Delta A)^2_{\psi_0}}\,.
\end{equation}
As
\begin{equation}
  \label{eq:67}
  |\sigma| =\sqrt{\gamma^2 +s^2} = \frac{(\Delta B)_{\psi_0}}{(\Delta A)_{\psi_0}}\,,
\end{equation}
the parameter $|\sigma|$ is a measure for the ``squeezing'' properties of the state $\psi_0$ with
respect to the two operators $A$ and $B$: it describes the ratio of the two uncertainties $
(\Delta B)_{\psi_0}$ and $(\Delta A)_{\psi_0}$.
\section{Minimal uncertainty states for $C,\,S$ and $L_{\dl}$}
As a first step let us determine functions $\psi_0$ which obey Eq.\ \eqref{eq:65} for the pair
$A=C = \cos \vp$ and $B=\tilde{L}_{\delta} = -i\partial_{\vp}$: The differential equation
\begin{equation}
  \label{eq:68}
 (\tilde{L}_{\delta} - \sigma\,\cos \vp)\psi_0(\vp) = (\langle \tilde{L}_{\delta}
\rangle_{\psi_0}-\sigma\,\langle C\rangle_{\psi_0})\psi_0(\vp)\,,~\sigma =\gamma-i\,s\,, 
\end{equation}
has the solutions
\begin{equation}
  \label{eq:69}
  \psi_0(\vp) = N\,e^{\ds i[(\tilde{l}_0-\sigma\,c_0)\vp + \sigma\,\sin \vp]}\,,\,\,N=
\mbox{const.}\,,
\end{equation}
where
\begin{equation}
  \label{eq:70}
  \tilde{l}_0 = \langle \tilde{L}_{\delta}  \rangle_{\psi_0}\,,\,\, c_0 = \langle C
 \rangle_{\psi_0}\,. 
\end{equation}
If $s\,c_0\neq 0$ then $\psi_0$ is not periodic or quasi-periodic (see \eqref{eq:42}), i.e.\
the solution \eqref{eq:69} would not belong to any Hilbert space $L^2(S^1,d\vp/2\pi,\delta)$.
As the commutator $[C,\tilde{L}_{\delta}]$ does not vanish we expect $s\neq 0$, according to
the second of the relations \eqref{eq:66}. So we assume
\begin{equation}
  \label{eq:71}
  c_0 \equiv \langle C \rangle_{\psi_0} =0\,.
\end{equation}
For
\begin{equation}
  \label{eq:72}
   \psi_0(\vp) =N\,e^{\ds i(\tilde{l}_0\,\vp + \sigma\,\sin \vp)}
\end{equation}
we have
\begin{equation}
  \label{eq:73}
  \psi_0(\vp+2\pi) = e^{\ds i2\pi\tilde{l}_0}\psi_0(\vp)\,.
\end{equation}
We can decompose the real number $\tilde{l}_0$ uniquely into an integer $n_0$ and a
 fractional part
$\delta_0$:
\begin{equation}
  \label{eq:74}
  \tilde{l}_0 = n_0+\delta_0\,,\,\, n_0\in \mathbb{Z}\,,\,\,\delta_0 \in [0,1)\,,  
\end{equation}
so that
\begin{equation}
  \label{eq:75}
   \psi_0(\vp+2\pi) =e^{\ds i2\pi\delta_0}\psi_0(\vp)\,.
\end{equation}
Thus, $\psi_0(\vp)$ is a possible element of the Hilbert space $L^2(S^1,d\vp/2\pi,\delta_0)$.
It yields the probability density
\begin{equation}
  \label{eq:76}
  |\psi_0(\vp)|^2 = |N|^2\,e^{\ds 2\,s\, \sin \vp}\,.
\end{equation}
For a given $s>0$ the density \eqref{eq:76} has its maximum at $\vp =\pi/2$ and its minimum
at $\vp = 3\pi/2$ for $\vp \in [0,2\pi)$. For $s<0$ the two are interchanged. 

 The normalization condition
\begin{equation}
  \label{eq:77}
 \int_{0}^{2\pi}\frac{d\vp}{2\pi}\,|\psi_0(\vp)|^2 =1 
\end{equation}
and assuming $N$ to be real and positive yield \cite{wat1}:
\begin{equation}
  \label{eq:78}
  \psi_0(\vp) =\frac{1}{\sqrt{I_0(2s)}}\,e^{\ds i(\tilde{l}_0\,\vp + \sigma\,\sin \vp)}\,. 
\end{equation}
(The modified Bessel function $I_0(2s)$ is always positive for real $s$ \cite{wat2}.)

We have
\begin{equation}
  \label{eq:79}
  \langle C \rangle_{\psi_0} = 0\,,\,\,\langle S \rangle_{\psi_0} = \frac{I_1(2s)}{I_0(2s)}\,,
\,\, \langle \tilde{L}_{\delta_0} \rangle_{\psi_0} = \tilde{l}_0\,.
\end{equation}
The second of the last Eqs.\ follows from
\begin{equation}
  \label{eq:80}
  \int_{0}^{2\pi}\frac{d\vp}{2\pi}\,\sin \vp\, e^{\ds 2s\,\sin \vp} = \frac{d}{d(2s)}I_0(2s) =
 I_1(2s)\,.
\end{equation}
($I_1(2s)$ is an odd function: $I_1(-2s)=-I_1(2s)$; furthermore  $|I_1(2s)|/I_0(2s)
< 1 $\,\cite{ka2}.)

As
\begin{equation}
  \label{eq:81}
 \int_{0}^{2\pi}\frac{d\vp}{2\pi}\,\sin^2 \vp e^{\ds 2s\,\sin \vp} = \frac{d}{d(2s)}I_1(2s) =
I_0(2s)-\frac{I_1(2s)}{2s}\,,  
\end{equation}
we have
\begin{equation}
  \label{eq:82}
  \langle S^2 \rangle_{\psi_0}= 1-\frac{ I_1(2s)}{2s\,I_0(2s)}\,,\,~~
 \langle C^2 \rangle_{\psi_0}= 1-\langle S^2 \rangle_{\psi_0} =
\frac{ I_1(2s)}{2s\,I_0(2s)}\,
\end{equation}
and
\begin{equation}
  \label{eq:83}
  \langle \tilde{L}_{\delta_0}^2\rangle_{\psi_0} = \tilde{l}_0^2 + |\sigma|^2\, \langle C^2 
\rangle_{\psi_0}=\tilde{l}_0^2 + |\sigma|^2\,\frac{I_1(2s)}{2s\,I_0(2s)}\,,
\end{equation}
so that
\begin{eqnarray}
  \label{eq:84}
 (\Delta C)^2_{\psi_0}& =& \frac{ I_1(2s)}{2s\,I_0(2s)}\,, \\
(\Delta S)^2_{\psi_0}& =& 1-\frac{ I_1(2s)}{2s\,I_0(2s)}-
 \frac{ I_1^2(2s)}{I_0^2(2s)}\,, \label{eq:319} \\
(\Delta \tilde{L}_{\delta_0})^2_{\psi_0} &=&|\sigma|^2\frac{ I_1(2s)}{2s\,I_0(2s)}\,.
\label{eq:320}
\end{eqnarray}
For the interpretation of the above and later formulae the following inequality is
 important: 
\begin{equation}
  \label{eq:97}
 0< \frac{I_1(2s)}{2s\,I_0(2s)} \leq \frac{1}{2}\,.
\end{equation}
It can be read off the series expansions of $I_1$ and $I_0$ (see Ref.\ \cite{wat2}). The equality
holds for $s=0$. For $|s| \to \infty$ the ratio \eqref{eq:97} tends to $0$.
 It follows from Eqs. \eqref{eq:84}, \eqref{eq:319} and \eqref{eq:320} that
\begin{equation}
  \label{eq:98}
  (\Delta C)^2_{\psi_0}+ (\Delta S)^2_{\psi_0}= 1-  \frac{ I_1^2(2s)}{I_0^2(2s)}\,,
\end{equation}
and
\begin{equation}
  \label{eq:321}
  (\Delta \tilde{L}_{\delta_0})^2_{\psi_0}/(\Delta C)^2_{\psi_0} = |\sigma|^2\,\,.
\end{equation}

For the correlation functions  \eqref{eq:64} we here have
\begin{equation}
  \label{eq:85}
  \langle S_{\psi_0}(C,\tilde{L}_{\delta_0})\rangle_{\psi_0} = \gamma\,\langle C^2
 \rangle_{\psi_0} =\gamma\,
\frac{ I_1(2s)}{2s\,I_0(2s)}\,,
\end{equation}
and
\begin{equation}
  \label{eq:90}
 \langle S_{\psi_0}(S,\tilde{L}_{\delta_0})\rangle_{\psi_0} = 0\,,\,\,~~~\langle
 S_{\psi_0}(C,S)\rangle_{\psi_0}=0\,.
\end{equation}

 From the first of the Eqs.\ \eqref{eq:52} we obtain
\begin{equation}
  \label{eq:86}
 \langle [C,\tilde{L}_{\dl_0}]\rangle_{\psi_0} = -i \langle S \rangle_{\psi_0} =
 -i\frac{I_1(2s)}{I_0(2s)}\,. 
\end{equation}

Collecting the corresponding formulae we can verify that the inequality \eqref{eq:63}
becomes an equality for $A=C,B=\tilde{L}_{\delta_0}$ and $\psi = \psi_0$ of Eq.\ \eqref{eq:78}:
\begin{equation}
  \label{eq:318}
   (\Delta C)^2_{\psi_0}\,(\Delta \tilde{L}_{\delta_0})^2_{\psi_0} =|\sigma|^2
\frac{ I_1^2(2s)}{4s^2\,I_0^2(2s)}= |\langle S_{\psi_0}(C,\tilde{L}_{\delta_0})
\rangle_{\psi_0}|^2+\frac{1}{4}| \langle [C,\tilde{L}_{\dl_0}]\rangle_{\psi_0}|^2 \,.
\end{equation}

Of interest are the limiting cases $s \to 0$ and $s \to + \infty$ for the 
parameter $s$:
 From \cite{ka3}
\begin{equation}
  \label{eq:87}
  \frac{I_1(2s)}{I_0(2s)} \to s\,(1-\frac{s^2}{2}) \mbox{ for } s \to 0\,;\,\,\,~~
 \frac{I_1(2s)}{I_0(2s)} \to 1-\frac{1}{4s} + O(s^{-2}) \mbox{ for } s \to +\infty\,
\end{equation}
it follows that  for $s \to 0$ and $\gamma$ fixed:
\begin{eqnarray}
  \label{eq:88}
  \langle S\rangle_{\psi_0} \to s\,,\,\,~~ (\Delta C)^2_{\psi_0} &\to& \frac{1}{2}
 + O(s^2)\,,\,\,
~~(\Delta S)^2_{\psi_0} \to \frac{1}{2} + O(s^2)\,,\\ (\Delta
 \tilde{L}_{\delta_0})^2_{\psi_0} &\to&
\frac{1}{2}\,|\sigma|^2 \to \frac{1}{2}\,\gamma^2 + O(s^2)\,. \nonumber
\end{eqnarray}  For $s \to \infty$:
\begin{eqnarray}
  \label{eq:89}
  \langle S\rangle_{\psi_0} \to 1-O(s^{-1})\,,\,\,~~ (\Delta C)^2_{\psi_0} &\to& \frac{1}{2s} -
 O(s^{-2})\,,\,\,~~
(\Delta S)^2_{\psi_0} \to \frac{1}{8s^2} + O(s^{-3})\,,\\ (\Delta
 \tilde{L}_{\delta_0})^2_{\psi_0} &\to&
\frac{s}{2} + O(1) \,. \nonumber
\end{eqnarray}

The 2 limiting cases \eqref{eq:88} and \eqref{eq:89} show rather obviously the complementarity
between the ``observables'' $C$ and $S$ on the one hand and $\tilde{L}_{\dl_0}$ on the other.
In the above discussion we considered the case of positive $s$. The case of negative $s$ can be
reduced to the positive one by observing that $I_0(-s)=I_0(s)\,,I_1(-s)=-I_1(s)$. 

Because of its quasi-periodicity \eqref{eq:75} the function $\psi_0$ may be expanded in terms
of the basis $e_{n,\delta_0}(\vp)$:
\begin{equation}
  \label{eq:91}
  \psi_0(\vp) = \sum_{n \in \mathbb{Z}} c_n\,e^{\ds i(\delta_0+n)\vp}\,,
\end{equation}
where \cite{wat3}
\begin{equation}
  \label{eq:92}
  c_n= (e_{n,\delta_0},\psi_0) = \frac{1}{\sqrt{I_0(2s)}}\int_0^{2\pi} \frac{d\vp}{2\pi}
 e^{\ds i[-(n-n_0)\vp+\sigma
 \sin \vp]} = J_{n-n_0}(\sigma)/\sqrt{I_0(2s)}\,.
\end{equation}
Here $J_n(z)$ is the Bessel function of order $n$. It has the property $J_{-n}(z)=(-1)^n J_n(z)$,
so that $|c_{-n}|^2 = |c_n|^2$.
Therefore the normalization condition
\begin{equation}
  \label{eq:93}
 \sum_{n=-\infty}^{n=+\infty} |c_n|^2 = 1 
\end{equation}
implies the ``sum rule''
\begin{equation}
  \label{eq:94}
 |J_0(\sigma)|^2 + 2\sum_{n=1}^{n=+\infty} |J_n(\sigma)|^2 = I_0(2s)\,,\,\,\sigma =\gamma -
i\,s \in 
\mathbb{C}\,. 
\end{equation}
This is a generalization of the well-known relations \cite{wat4}
\begin{equation}
  \label{eq:99}
   \sum_{n=-\infty}^{n=+\infty} J_n(z)\,J_{-n}(z)= J_0(2z)
\end{equation}
and
\begin{equation}
  \label{eq:95}
  J_0^2(x) + 2\,\sum_{n=1}^{n= +\infty} J_n^2(x) =1\,,\,\,x \in \mathbb{R}\,.
\end{equation}
(One has $I_0(2s=0)=1$.) 

The relation \eqref{eq:94} implies
  \begin{equation}
    \label{eq:96}
    |c_0|^2 \leq 1\,\,,~~|c_n|^2 \leq 1/2\, \mbox{ for }\,n=1,2,\ldots\,\,.
  \end{equation}.

The parameters $\gamma$ and $s$ obviously characterize properties of the probability
distribution associated with the wave function \eqref{eq:78}: 

 The parameter $s$ corresponds
to the parameter $a>0$ in the Gaussian wave packet
\begin{equation}
  \label{eq:100}
  \psi_G(x) = (\frac{2a}{\pi})^{1/4}\,e^{\ds -a\,x^2}
\end{equation}
and determines the width of the distribution. The correlation \eqref{eq:64}, and therefore
 $\gamma$, vanish for the wave
function \eqref{eq:100} with $A=Q,B=P$.

 In our case the parameter $\gamma$ here  describes
 the correlations \eqref{eq:85}. A combination of $\gamma$
 and $s$, namely $|\sigma|^2 = \gamma^2 + s^2$ determines the squeezing properties of the
distribution [see Eq.\ \eqref{eq:321}]. If $\gamma$ vanishes then $s$ alone
characterizes the distribution and its squeezing properties.

The real number $l_0 = \hbar\, \tilde{l}_0$ is the expectation value of the operator
 $L_{\delta_0}$ with respect to the wave function \eqref{eq:78}.
It corresponds to the classical orbital angular momentum $p_{\vp}$. 

The wave function \eqref{eq:78} does not contain a  parameter corresponding to a classical
angle $\alpha$  which represents the angle $\vp$ of the classical phase space \eqref{eq:12}.
This can be taken care of by the replacement
\begin{equation}
  \label{eq:101}
  \psi_0(\vp) \to \psi_{\alpha,\tilde{l}}(\vp) = \psi_0(\vp-\alpha) = 
\frac{1}{\sqrt{I_0(2s)}}\,e^{\ds i[\tilde{l}\,(\vp-\alpha) + \sigma\,\sin (\vp-\alpha)]}\,.
\end{equation}
(We now drop the index 0 of the number $\tilde{l}_0$.)

The expectation values of $C,\,S$ and $\tilde{L}_{\delta}$ with respect to the wave function
\eqref{eq:101} may be reduced to the previous ones by observing that for any periodic function
$f(\vp+2\pi) = f(\vp)$ we have
\begin{equation}
  \label{eq:102}
  \int_c^{2\pi +c}d\vp\, f(\vp) = \int_0^{2\pi}d\vp\,f(\vp)\,.
\end{equation}
Thus, e.g.\ we obtain
\begin{equation}
  \label{eq:103}
  \langle C \rangle_{\alpha,\tilde{l}} \equiv (\psi_{\alpha,\tilde{l}},C\,\psi_{\alpha,
\tilde{l}})= \int_0^{2\pi}\frac{d \vp}{2\pi}\,\cos \vp\, e^{\ds 2\,s\,\sin (\vp-\alpha)} =
\int_0^{2\pi}\frac{d \vp}{2\pi}\,\cos (\vp+\alpha)\, e^{\ds 2\,s\,\sin \vp}\,.
\end{equation}
Observing that $\cos(\vp+\alpha) = \cos \alpha \,\cos \vp - \sin \alpha \, \sin \vp$ we get
from the relations \eqref{eq:79}:
\begin{equation}
  \label{eq:104}
 \langle C \rangle_{\alpha,\tilde{l}} = -\sin \alpha\,\frac{I_1(2s)}{I_0(2s)}\,.
\end{equation}

In the same way we have
\begin{eqnarray}
  \label{eq:105}
   \langle S \rangle_{\alpha,\tilde{l}}& =& \cos \alpha\,\frac{I_1(2s)}{I_0(2s)}\,, \\
 \langle C \rangle_{\alpha,\tilde{l}}^2 + \langle S \rangle_{\alpha,\tilde{l}}^2 &=&
\frac{I_1^2(2s)}{I_0^2(2s)}\, \nonumber \\
 \langle \tilde{L}_{\delta} \rangle_{\alpha,\tilde{l}}&=& \tilde{l} = n+\delta \,,\,\dl \in
[0,1)\,,\label{eq:322} \\
  \label{eq:106}
  \langle C^2 \rangle_{\alpha,\tilde{l}}& =& \cos 2 \alpha\,\frac{I_1(2s)}{2s\,I_0(2s)}
+ \sin^2 \alpha\,,\\ \langle S^2 \rangle_{\alpha,\tilde{l}}&=&-\cos 2 \alpha\,\frac{I_1(2s)}{
2s\,I_0(2s)}+ \cos^2 \alpha\,,\label{eq:110} \\
  \label{eq:107}
 \langle \tilde{L}_{\delta}^2\rangle_{\alpha,\tilde{l}}& =&
 \tilde{l}^2 + |\sigma|^2\frac{I_1(2s)}{2s\,I_0(2s)}\,, 
\end{eqnarray}
from which we get
\begin{eqnarray}
  \label{eq:108}
  (\Delta C)^2_{\alpha,\tilde{l}}& =& \cos 2 \alpha\,
\frac{I_1(2s)}{2s\,I_0(2s)}\,
+\, \sin^2 \alpha\,\left(1-\frac{I_1^2(2s)}{I^2_0(2s)}\right)\,,\\
(\Delta S)^2_{\alpha,\tilde{l}}&=&-\cos 2 \alpha\,\frac{I_1(2s)}{
2s\,I_0(2s)}+ \cos^2 \alpha\left (1-\frac{I^2_1(2s)}{I_0^2(2s)}\right)\,, \label{eq:111}
\\(\Delta \tilde{L}_{\delta})^2_{
\alpha,\tilde{l}} &=&|\sigma|^2\frac{ I_1(2s)}{2s\,I_0(2s)}\,. \label{eq:112}
\end{eqnarray}
Notice that the sums of the expressions \eqref{eq:106} and \eqref{eq:110}, and \eqref{eq:108}
and \eqref{eq:111} respectively, are independent of $\alpha$\,!

Furthermore, for the correlation function \eqref{eq:64} we here have
\begin{eqnarray}
  \label{eq:109}
  \langle S_{\alpha,\tilde{l}}(C,\tilde{L}_{\delta})\rangle_{\alpha,\tilde{l}}& =& \gamma\,\cos
\alpha\,\frac{I_1(2s)}{2s\,I_0(2s)}\,,\\ \langle S_{\alpha,\tilde{l}}(S,\tilde{L}_{\delta})
\rangle_{\alpha,\tilde{l}} &=& \gamma\,\sin
\alpha\,\frac{I_1(2s)}{2s\,I_0(2s)}\,, \label{eq:113} \\ \langle S_{\alpha,
\tilde{l}}(C,S)\rangle_{\alpha,
\tilde{l}}& =& \frac{1}{2}\sin 2\alpha \left(\frac{I_1(2s)}{2s\,I_0(2s)} +
 \frac{I_1^2(2s)}{I_0^2(2s)}-\frac{1}{2}\right)\,. \label{eq:114}
\end{eqnarray}
For $\alpha \neq 0$ the wave functions \eqref{eq:101} no longer minimize the uncertainty relation
\eqref{eq:63}:
\\ From Eqs.\ \eqref{eq:108} and \eqref{eq:112} we have
\begin{eqnarray}
  \label{eq:115}
  (\Delta C)^2_{\alpha,\tilde{l}}\,(\Delta \tilde{L}_{\delta})^2_{
\alpha,\tilde{l}} &=& |\sigma|^2\frac{ I_1(2s)}{2s\,I_0(2s)}\,\left[\cos 2 \alpha\,
\frac{I_1(2s)}{2s\,I_0(2s)}\,
+\, \sin^2 \alpha\,\left(1-\frac{I_1^2(2s)}{I^2_0(2s)}\right) \right] \\ &=&
|\sigma|^2\,\cos^2\alpha\,\frac{I_1^2(2s)}{4s^2I_0^2(2s)}+|\sigma|^2\,\sin^2\alpha\, 
\frac{I_1(2s)}{2s\,I_0(2s)}\left[1-\frac{I_1^2(2s)}{I_0^2(2s)}-\frac{I_1(2s)}{2s\,I_0(2s)} 
\right]\,,\nonumber
\end{eqnarray}
whereas the Eqs.\ \eqref{eq:109} and \eqref{eq:105} give
\begin{equation}
  \label{eq:116}
  |\langle S_{\alpha,\tilde{l}}(C,\tilde{L}_{\delta})\rangle_{\alpha,\tilde{l}}|^2 + \frac{1}{4}
|\langle S\rangle_{\alpha,\tilde{l}}|^2 =|\sigma|^2\,\cos^2\alpha\frac{I_1^2(2s)}{4s^2
I_0^2(2s)}\,.
\end{equation}

The function
\begin{equation}
  \label{eq:314}
  g(x)=1-\frac{I_1^2(x)}{I_0^2(x)}-\frac{I_1(x)}{x\,I_0(x)}=  g(-x)\,,~~x=2s\,,
\end{equation}
which appears in Eqs.\ \eqref{eq:319},\,\eqref{eq:114}\,,\,[ here as $0.5-g(x)$]\,,\,  and
 \eqref{eq:115}, varies between 
$0.5$ and $0$ if $|s|$ varies between $0$ and $\infty$. Numerical examples \cite{abr} for
several functions appearing in the present chapter:
\begin{center} \begin{tabular}{|c|c|c|c|}\hline $x$&$I_1(x)/I_0(x)$&$I_1(x)/[x\,I_0(x)]$&
 $g(x)$ \\ \hline \hline
0&0&0.5 & 0.5 \\ \hline 0.1& 0.0499& 0.4994 & 0.4981 \\ \hline 0.5 & 0.2425&0.4850 & 0.4562
 \\ \hline
1 & 0.4464& 0.4464 & 0.3543\\ \hline 2& 0.6977&0.3489 & 0.1644 \\ \hline 5&0.8934& 0.1787
 & 2.32$\cdot 10^{-2}$ \\
\hline
 10 & 0.9486& 9.47 $\cdot 10^{-2}$& 5.29$\cdot 10^{-3}$ \\ \hline 50& 0.9900& 1.95$\cdot 10^{-2}$
& 1.99$\cdot 10^{-4}$ \\ \hline
100&0.9950&9.95$\cdot 10^{-3}$ &4.60$\cdot 10^{-5}$ \\ \hline \end{tabular} \end{center}
Asymptotically we have for large $|x|$ \cite{ka3}:
\begin{equation}
  \label{eq:315}
  g(x) \asymp \frac{1}{2x^2} + O(x^{-3})~\text{for}~|x| \to \infty\,.
\end{equation}

The numerical examples show how one can influence the different expectation values and mean
square deviations by a suitable choice of the parameter $s\,$!

For the special value $\alpha =\pi/2$ we have $\sin(\vp-\pi/2)=-\cos\vp$ and the wave
 function \eqref{eq:101}
 becomes the minimal uncertainty wave function 
 \begin{equation}
   \label{eq:117}
 \frac{1}{\sqrt{I_0(2s)}}\,e^{\ds i[\tilde{l}\,(\vp-\pi/2) - \sigma\,\cos \vp]}  
 \end{equation}
for the product $ (\Delta S)^2_{\alpha=\pi/2,\tilde{l}}\,(\Delta \tilde{L}_{\delta})^2_{
\alpha=\pi/2,\tilde{l}}$, if we replace $C=\cos\vp$ in Eq.\ \eqref{eq:68} by $S=\sin\vp$:
 According to Eqs.\ \eqref{eq:111}, \eqref{eq:112}, \eqref{eq:113} and \eqref{eq:104} we have now
\begin{equation}
  \label{eq:118} (\Delta S)^2_{\alpha=\pi/2,\tilde{l}}\,(\Delta \tilde{L}_{\delta})^2_{
\alpha=\pi/2,\tilde{l}} =
  |\langle S_{\alpha=\pi/2,\tilde{l}}(S,\tilde{L}_{\delta})\rangle_{\alpha=\pi/2,\tilde{l}}|^2
 + \frac{1}{4}
|\langle C\rangle_{\alpha=\pi/2,\tilde{l}}|^2 =|\sigma|^2\,\frac{I_1^2(2s)}{4s^2I_0^2(2s)}\,.
\end{equation}
As $\cos \vp = 1- \vp^2/2 + O(\vp^4)$ the functions \eqref{eq:117} are locally Gaussian ones for 
$\vp^2 \ll 1$ and $s <0$, but, of course, not globally! In that local limit the 2nd of
 Eqs.\ \eqref{eq:52}
 yields $[L_{\dl},\vp]= -i\,\hbar + O(\vp^2)$.

Notice that the r.h.\ side of Eq.\ \eqref{eq:116} vanishes now, but the r.h.\ side of Eq.
\eqref{eq:115} does not!

Instead of the coefficients \eqref{eq:92} we now get for the functions \eqref{eq:101}
\begin{eqnarray}
  \label{eq:119}
  c_m=(e_{m,\delta},\psi_{\alpha,\tilde{l}=n+\dl})&=&\frac{e^{\ds-i(m+\delta)\alpha}}{
\sqrt{I_0(2s)}}
\int_0^{2\pi} \frac{d\vp}{2\pi}
 e^{\ds i[-(m-n)\vp+\sigma
 \sin \vp]}\\& =&\frac{e^{\ds-i(m+\delta)\alpha}}{\sqrt{I_0(2s)}} J_{m-n}(\sigma)\,,~ 
  m \in \mathbb{Z}\,. \nonumber 
\end{eqnarray}
The last result and the relation \eqref{eq:94} can be used to show that the states
 \eqref{eq:101} form a complete set:
\begin{eqnarray}
  \label{eq:120}
  \int_0^{2\pi}\frac{d\alpha}{2\pi}\sum_{n=-\infty}^{n=+\infty}(e_{m_1,\delta},\psi_{\alpha,
\,n+\delta})\,(\psi_{\alpha,\,n+\delta},e_{m_2,\delta}) &=& \delta_{m_1m_2} \frac{1}{I_0(2s)}\,
\sum_{n=-\infty}^{n=+\infty} J^*_{m_2-n}(\sigma)\,J_{m_1-n}(\sigma)~~~~\\
&=& 1 \mbox{ for } m_2 =m_1,\nonumber \\& =& 0 \mbox{ for } m_2 \neq m_1\,. \nonumber
\end{eqnarray}
Two different states \eqref{eq:101} are not orthogonal:
\begin{eqnarray}
  \label{eq:121}
  (\psi_{\alpha_2,\tilde{l}_2}, \psi_{\alpha_1,\tilde{l}_1})& =& \frac{e^{\ds
 i(\alpha_2-\alpha_1)
(\tilde{l}_1+\tilde{l}_2)/2}}{I_0(2s)}\left[\frac{(\gamma\,\sin[\frac{1}{2}(\alpha_1-
\alpha_2)]-s\,
\cos[\frac{1}{2}(\alpha_1-\alpha_2)])}{(\gamma\,\sin[\frac{1}{2}(\alpha_1-\alpha_2)]+s\,
\cos[\frac{1}{2}(\alpha_1-\alpha_2)])}\right]^{\ds(\tilde{l}_1-\tilde{l}_2)/2} ~~~~~~ \\
&& \times \,I_{\tilde{l}_1-
\tilde{l}_2}\left(2\sqrt{s^2\,
\cos^2[\frac{1}{2}(\alpha_1-\alpha_2)]-\gamma^2\,\sin^2[\frac{1}{2}(\alpha_1-\alpha_2)]}\right)
\,. \nonumber
\end{eqnarray}
The matrix element \eqref{eq:121} may be calculated as follows: First use the relations
\begin{eqnarray}
  \label{eq:122}
  \sin(\vp-\alpha_1)+\sin(\vp-\alpha_2) &=& 2\,\cos[\frac{1}{2}(\alpha_1-\alpha_2)]\,
\sin[\vp-\frac{1}{2}(\alpha_1+\alpha_2)]\,,\\
\sin(\vp-\alpha_1)-\sin(\vp-\alpha_2) &=& 2\,\sin[\frac{1}{2}(\alpha_1-\alpha_2)]\,
\cos[\vp-\frac{1}{2}(\alpha_1+\alpha_2)] \nonumber
\end{eqnarray}
 for the integrand of
\[ \int_0^{2\pi}\frac{d\vp}{2\pi}\psi^*_{\alpha_2,\tilde{l}_2}\,\psi_{\alpha_1,\tilde{l}_1}\,.
\]The resulting integral can be evaluated by using integral tables \cite{hof}.

Parts of the wave functions \eqref{eq:101} have been discussed previously:

De Bi\`{e}vre \cite{bie} and later Torresani \cite{torr} considered the functions
\begin{equation}
  \label{eq:123}
  \chi_{\alpha,\gamma}(\vp) = e^{\ds i\,\gamma\,\sin(\vp-\alpha)}\,,~\gamma \in \mathbb{R}\,,
\end{equation}
 and the associated integral transforms
\begin{equation}
  \label{eq:124}
  \tilde{\eta}(\alpha,\gamma)  = \int_{-\pi/2}^{+\pi/2}d\vp\,\chi_{\alpha,\gamma}\,\eta(\vp)\,.
\end{equation}
Their approach was motivated by the problem that the Perelomov construction of coherent states
for Lie groups \cite{pere} does not work for the group $E(2)$ because the irreducible unitary
 representations \eqref{eq:24}-\eqref{eq:25} are not integrable in the following sense:
Let us put $\delta =0$ and combine the two formulae into one (with $\hbar=1$):
\begin{equation}
  \label{eq:125}
  [U^{\rho}(\alpha,a,b)\psi](\vp) = e^{\ds -i\,\rho\,(a\,\cos\vp+b\sin\vp)}\psi(\vp-\alpha)\,.
\end{equation}
Then one can show that
\begin{equation}
  \label{eq:126}
  \int_{E(2)}d\alpha\, da\, db |(\psi,U^{\rho}(\alpha,a,b)\psi)|^2 = \infty\,,~ \psi(\vp) \in
 L^2(S^1,d\vp/2\pi)\,.
\end{equation}
Note that  the function \eqref{eq:123} is closely related to the transformation \eqref{eq:125}
 if one
puts $a=0$ and $b=-\gamma$.

The ``coherent'' states \eqref{eq:123} are unsatisfactory for the following reasons:

We have already described above that the parameter $\gamma$ characterizes the distribution,
not the expectation value of a physical observable (one has $(\chi_{\alpha,\gamma},L\chi_{\alpha,
\gamma})=0$\,!). In addition there are problems with the completeness relation which here 
requires
that the function $\eta(\vp)$ in \eqref{eq:124} has to obey - among others - the condition
\begin{equation}
  \label{eq:127}
  \int_{-\pi/2}^{\pi/2}d\vp\, \frac{|\eta(\vp)|^2}{\cos\vp} < \infty\,,
\end{equation}
which means that $\eta(\vp)$ should vanish sufficiently enough at $\vp = \pm \pi/2$\,.
The problem may be exhibited heuristically in the following way: We have (see Eq.\
 \eqref{eq:119}) \begin{equation}
  \label{eq:128}
  c_m=(e_m,\chi_{\alpha,\gamma})=\int_0^{2\pi}\frac{d\vp}{2\pi}\,e^{\ds i(\gamma\,
\sin(\vp-\alpha)
 -im\vp)} =e^{\ds -im\alpha}\,J_n(\gamma)\,,
\end{equation}
from which it follows that
\begin{equation}
  \label{eq:129}
  \int_{-\infty}^{+\infty}d\gamma \int_0^{2\pi}\frac{d\alpha}{2\pi}\, (e_{m_1},
\chi_{\alpha,\gamma})
\,(\chi_{\alpha,\gamma},e_{m_2}) = \delta_{m_1m_2}\int_{-\infty}^{+\infty}d\gamma\, J_{m_1}^2(
\gamma)=2\,  \delta_{m_1m_2}\int_0^{+\infty}d\gamma\, J_{m_1}^2(
\gamma)\,.
\end{equation}
As \cite{wat5}
\begin{equation}
  \label{eq:130}
  J_n^2(\gamma) = \frac{1}{\pi}\int_0^{\pi}d\vp\,J_0(2\gamma\,\sin\vp)\,\cos(2n\vp)\,,~~
\int_0^{\infty}d\gamma\,e^{\ds -\epsilon\,\gamma}J_0(2\sin\vp\,\gamma) = \frac{1}{\sqrt{
\epsilon^2+4\sin^2\vp}}\,,
\end{equation}
we have
\begin{equation}
  \label{eq:131}
  \int_0^{\infty}d\gamma\,e^{-\epsilon\,\gamma}J_n^2(\gamma) =\frac{1}{\pi}\int_0^{\pi}
d\vp\,\frac{\cos(2n\vp)}{\sqrt{\epsilon^2+4\sin^2\vp}}\,.
\end{equation}
Taking the limit $\epsilon \to 0$ we see that the integral \eqref{eq:129} diverges.

Isham and Klauder \cite{kl} avoided the difficulties \eqref{eq:126} by introducing an additional
averaging over the parameter $\rho$, i.e.\ by averaging over different irreducible unitary
representations. Such averaging modifies e.g.\ the integral transform \eqref{eq:124}. We
 have seen above that such a procedure is not necessary if one allows for quasi-OAM.

Kowalski and Rembieli\'{n}ski  mention the states \eqref{eq:101}
 (with $\gamma =0$) in the
introduction of their paper Ref.\ \cite{kow1}, but discard them, because they allow only
 for representations
with $\dl=0$, i.e.\ $\tl$ would have to be an integer!

\section{Generating coherent states on the circle by means of the Weil-Zak
 transform}
There is an elegant way of generating coherent states on the circle from those well-known ones
of the harmonic oscillator. The method makes use of a transform discussed by the mathematician
Weil \cite{weil} and independently by the physicist Zak \cite{zak}. In the present context it
has been introduced and employed by De Bi\`{e}vre and Gonz\'{a}lez \cite{gon1} and Gonz\'{a}lez
and del Olmo \cite{gon2}. It leads automatically to the introduction of fractional orbital
 angular momenta $\delta$! The basic idea may be sketched as follows \cite{reed4}:

Consider a function $f(\xi) \in L^2(\mathbb{R},d\xi)$, then one can define a
 function $f^{(\delta)} (\vp)$ on the unit circle by 
\begin{equation}
  \label{eq:132}
  f^{(\delta)}(\vp) = \sum_{n \in \mathbb{Z}}e^{\ds -i2\pi\delta n}f(\vp + 2\pi n)\,,~~
\vp \in [0,2\pi)\,,~~\delta \in [0,1)\,.
\end{equation}
The function  $f^{(\delta)}(\vp)$ has the following properties:
\begin{equation}
  \label{eq:133}
  f^{(\delta)}(\vp +2\pi)=e^{\ds i2\pi\delta}f^{(\delta)}(\vp)\,,
\end{equation}
with $\tilde{\delta}=2\pi\delta$ one has
\begin{equation}
  \label{eq:134}
  \int_0^{2\pi}\frac{d\tilde{\delta}}{2\pi}\int_0^{2\pi}d\vp\,|f^{(\delta)}(\vp)|^2
= \int_0^{2\pi}d\vp\, \sum_{n \in \mathbb{Z}}|f(\vp+2\pi\,n)|^2 = \int_{-\infty}^{+\infty}
d\xi \,|f(\xi)|^2\,. 
\end{equation}
The inverse of the transform \eqref{eq:132} is given by
\begin{equation}
  \label{eq:135}
  f(\vp+2\pi n) = \int_0^{2\pi}\frac{d\tilde{\delta}}{2\pi}\,e^{\ds i n \tilde{\delta}}
f^{(\delta)}(\vp)
=\int_0^{2\pi}\frac{d\tilde{\delta}}{2\pi}\,e^{\ds i n\tilde{\delta}}\,\sum_{m \in
 \mathbb{Z}} e^{\ds -im\tilde{\delta}}f(\vp + 2\pi m)\,.
\end{equation}

The normalized coherent states on $L^2(\mathbb{R},dx)$ associated with the harmonic oscillator
are \cite{mand}
\begin{equation}
  \label{eq:136}
  u_{\alpha}(x)= \frac{1}{(\pi \lambda^2_0)^{1/4}}e^{\ds -(|\alpha|^2+\alpha^2)/2}\,
e^{\ds -(x/\lambda_0)^2/2 +\sqrt{2}\alpha x/\lambda_0}\,,~\lambda_0 =\sqrt{\frac{\hbar}{m\,\omega}}\,,
\end{equation}
where the complex numbers $\alpha =(q/\lambda_0+i\lambda_0p/\hbar)/\sqrt{2}$ are
 eigenvalues of the
annihilation operator
\begin{equation}
  \label{eq:137}
  a =\frac{1}{\sqrt{2}}(Q/\lambda_0+i\lambda_0 P/\hbar)\,.
\end{equation}
For the following it is convenient to introduce dimensionless quantities 
\begin{equation}
  \label{eq:138}
  \xi =x/\lambda_0\,,~~\tilde{q}= q/\lambda_0\,,~~\tilde{p}=\lambda_0p/\hbar\,,
~~z=\sqrt{2}\alpha = \tilde{q} +i\,\tilde{p}\,.
\end{equation}
We then have on $L^2(\mathbb{R},d\xi)$
\begin{equation}
  \label{eq:139}
  u_z(\xi) = (\pi)^{-1/4}e^{\ds-(|z|^2+z^2)/4}\,e^{\ds- \xi^2/2+z\xi}\,,
\end{equation}
where the factor holomorphic in $z$,
\begin{equation}
  \label{eq:324}
  e^{\ds -z^2/4-z\xi}= \sum_{n=0}^{\infty} \frac{z^n}{2^n\,k!}\,H_n(\xi)\,,
\end{equation}
is a generating function for the orthogonal Hermite polynomials $H_n(\xi)$.

For the ensuing discussions it is instructive to introduce a dimensionless
 parameter $\epsilon >0$ which has the value $\epsilon =1$ for the quantum theory
and which characterizes the classical limit $\hbar \to 0$ as $\epsilon \to 0$. This
 can be done
 - compare \eqref{eq:20} and \eqref{eq:136} - by replacing the wave function \eqref{eq:139} by
\begin{equation}
  \label{eq:155}
 u_z^{(\epsilon)}(\xi) = (\epsilon\pi)^{-1/4}e^{\ds-(|z|^2+z^2)/(4\epsilon)}\,e^{\ds- 
\xi^2/(2\epsilon)+z\xi/\epsilon}\,,
\end{equation}  
with the properties
\begin{eqnarray}
  \label{eq:156}
  \int_{\mathbb{R}}d\xi\,{u_z^{(\epsilon)}}^{*}(\xi),u_z^{(\epsilon)}(\xi) &=&1\,,\\
\int_{\mathbb{R}^2}\frac{d\tilde{q}\,d\tilde{p}}{2\pi\epsilon} {u_z^{(\epsilon)}}^{*}(\xi_1)\,
u_z^{(\epsilon)}(\xi_2) &=& \delta(\xi_1-\xi_2)\,. \label{eq:157}
\end{eqnarray}
The relation \eqref{eq:157} represents the completeness of the functions \eqref{eq:155}
(here $\dl(\xi)$ stands for the usual $\dl$-function!)

If we define the dimensionless operators
\begin{equation}
  \label{eq:232}
  \tilde{Q} =\xi\,,~~\tilde{P} =\frac{1}{i}\partial_{\xi}\,,
\end{equation}
we have the expectation values
\begin{equation}
  \label{eq:233}
\langle \tilde{Q} \rangle_{z,\epsilon} \equiv 
 \int_{\mathbb{R}}d\xi\,{u_z^{(\epsilon)}}^{*}(\xi)\,\tilde{Q}\,u_z^{(\epsilon)}
(\xi)=\tilde{q}\,,
~~\langle \tilde{P} \rangle_{z,\epsilon} = \tilde{p}/\epsilon\,,
\end{equation}
so that
\begin{equation}
  \label{eq:234}
  \tilde{a}^{(\epsilon)}u_z^{(\epsilon)}=z\,u_z^{(\epsilon)}\,,~~ \tilde{a}^{(\epsilon)} =
\tilde{Q}+i\epsilon\tilde{P}\,.
\end{equation}
Eqs.\ \eqref{eq:233}, \eqref{eq:234} and
\begin{equation}
  \label{eq:323}
  ((\Delta\tilde{Q})_{z,\epsilon})^2=\frac{\epsilon}{2}\,,~~
\epsilon^2 ((\Delta\tilde{P})_{z,\epsilon})^2=\frac{\epsilon}{2}\,,
\end{equation}
show that for the classical limit $\epsilon \to 0$ of matrix elements the
product $\epsilon\,\tilde{P}$ should be kept fixed, because $\tilde{p}$ as defined in Eq.\
\eqref{eq:138} represents the classical momentum and $\epsilon$ stands for Planck's constant
made dimensionless.

In case one wants to discuss matrix elements of the (dimensionless) operator $\tilde{P}$, then
$\epsilon$ serves as a squeezing parameter (divide the second of Eqs.\ \eqref{eq:323} by 
$\epsilon^2$!)

 Applying the mapping \eqref{eq:132} to the states \eqref{eq:155} yields
\begin{align}\label{eq:140}
  u^{(\epsilon,\delta)}_z(\vp)&= (\epsilon\pi)^{-1/4}\,e^{\ds -
 (|z|^2+z^2)/(4\epsilon)}\,
\sum_{n\in \mathbb{Z}}
e^{\ds -i2\pi n\dl}\,e^{\ds -(\vp+2\pi n)^2/(2\epsilon) + z(\vp+2\pi n)/\epsilon} \\
& = (\epsilon\pi)^{-1/4}\, e^{\ds -
 [(|z|^2-z^2)/(4\epsilon) +
(\vp -z)^2/(2\epsilon)]}\,
\vt_3[i\pi(\vp-z+i\epsilon\delta)/\epsilon ,\,e^{\ds-2\pi^2/\epsilon}\,]\,, \nonumber 
\end{align}
where
\begin{eqnarray}
  \label{eq:141}
  \vt_3 (\zeta,q=e^{\ds i\pi \tau})\equiv \vt_3(\zeta|\tau)& =& \sum_{n \in \mathbb{Z}}
q^{\ds n^2}\,
e^{\ds 2n i \zeta}=1+\sum_{n=1}^{\infty}q^{\ds n^2}\cos 2n\zeta\,,\\
&& \Im( \tau) > 0\,,~\vt_3(-\zeta,q)=\vt_3(\zeta,q)\,, \nonumber
\end{eqnarray}
is the third of Jacobi's $\vt$-functions which is an entire (holomorphic) function of $\zeta$
 (see the literature quoted in Appendix C).

 In Eq.\ \eqref{eq:140} we have $\tau = 2i\pi/\epsilon $. For real $q$ the function $\vt_3$ is
 real-valued for real and ima\-ginary arguments $\zeta$. It has its zeros at the points
 $ \zeta_0 = (m+1/2)\pi +
(n+1/2)\pi \tau,\, m,n \in \mathbb{Z}$. If $\tau$ is purely imaginary, there are no zeros 
on the real or imaginary axis and $\vt_3$ is positive there. 

Using Jacobi's famous identity
\begin{equation}
  \label{eq:142}
  \vt_3(\zeta|\tau) = (-i\tau)^{-1/2}\,e^{\ds \zeta^2/(i\pi \tau)}\vt_3(\zeta/\tau|-1/\tau),
\end{equation}
we can express Eq.\ \eqref{eq:140} as
\begin{equation}
  \label{eq:143}
  u_z^{(\epsilon,\delta)}(\vp) = \frac{1}{\sqrt{2\pi}}\left(\frac{\epsilon}{\pi}
\right)^{1/4}
\, e^{\ds 
 -(|z|^2-z^2)/(4\epsilon)}\,
e^{\ds[ 
i(\vp- z)\delta
-\epsilon\delta^2/2]}\,\vt_3[(\vp- z+i\epsilon\delta)/2,\,q=e^{\ds -\epsilon/2}]\,.
\end{equation}
As $\vt_3(\zeta,q)$ has the period $\pi$ in $\zeta$ it follows immediately from Eq.\ 
\eqref{eq:143} that
\begin{equation}
  \label{eq:144}
  u_z^{(\epsilon,\delta)}(\vp+2\pi)= e^{\ds i 2\pi\delta}\,  u_z^{(\epsilon,\delta)}(\vp)\,. 
\end{equation}
We now interpret the complex number $z$ as
\begin{equation}
  \label{eq:145}
  z=\theta+i\,\tilde{l}\,,~~\theta = \mathbb{R} \bmod{2\pi}\,,~\tilde{l} \in \mathbb{R}\,,
\end{equation}
so that
\begin{equation}
  \label{eq:146}
  |z|^2-z^2 = -2iz\tilde{l} =2(\tilde{l}^2 -i\theta\tilde{l}) \,.
\end{equation}
The scalar product
\begin{equation}
  \label{eq:147}
 (e_{m,\delta},u_z^{(\epsilon,\delta)}) = 
 \frac{1}{\sqrt{2\pi}}\left(\frac{\epsilon}{\pi}\right)^{1/4}\,
e^{\ds-[\tilde{l}-\epsilon(m+\delta)]^2/(2\epsilon)}
\,e^{\ds i\theta\,[\tilde{l}/(2\epsilon)-(m+\delta)]}\,
\end{equation}
 yields the completeness relation
\begin{align}  \label{eq:148}
& \int_{-\infty}^{+\infty}\frac{d\tilde{l}}{\epsilon}\,\int_0^{2\pi}
d\theta\,(e_{m_1,\delta},
u^{(\epsilon,\delta)}_z)\,
(u^{(\epsilon,\delta)}_z,e_{m_2,\delta})=  \\ &=
 \frac{1}{2\pi}\,\frac{1}{\sqrt{\epsilon\pi}}\, \int_{-\infty}^{+\infty}
d\tl\,
\int_0^{2\pi}d\theta\,e^{\ds i(m_2-m_1)\theta}\,e^{\ds- [(\tilde{l}- \epsilon(m_1 +
\delta
 ))^2   
+(\tilde{l}-\epsilon(m_2+\delta))^2)]/(2\epsilon)}=   \nonumber \\ & = \delta_{m_1m_2}\,,    
\nonumber \end{align}
where 
\begin{equation}
  \label{eq:158}
\int_{S^1\times \mathbb{R}}  d\mu(\theta,\tl) \equiv\int_{S^1}\frac{d\theta}{2\pi}\,
\int_{\mathbb{R}}\frac{d\tl}{\sqrt{\epsilon\pi}}\,e^{\ds -(\tl-c)^2/\epsilon}
=1\,,~c= \text{const.}\,,
\end{equation}
has been used.

For the function \eqref{eq:143} we get the scalar product
\begin{eqnarray}
  \label{eq:150}
  (u_z^{(\epsilon,\delta)},u_z^{(\epsilon,\delta)}) &=& \frac{1}{2\pi}\left(
\frac{\epsilon}{\pi}\right)^{1/2}\,
e^{\ds-(\tilde{l}-\epsilon\delta)^2/\epsilon}\,
\vt_3[i(\tilde{l}-\epsilon\delta),\,q=e^{\ds -\epsilon}\,] \\
&=& \frac{1}{2\pi}\vt_3[\pi(\tilde{l}-\epsilon\delta)/\epsilon,\,q=e^{\ds -\pi^2/\epsilon}\,]\,.
 \label{eq:151} 
\end{eqnarray}
The equality \eqref{eq:151} again is a consequence of the identity \eqref{eq:142}.
Thus, we have the normalized coherent states
\begin{equation}
  \label{eq:159}
\hat{u}_z^{(\epsilon,\delta)}(\vp) = C_z^{(\epsilon,\dl)}\,u_z^{(\epsilon,\delta)}(\vp)\,,~
C_z^{(\epsilon,\dl)}=\left(\frac{2\pi}{
\vt_3[\pi(\tl-\epsilon\dl)/\epsilon,\,q=e^{\ds-\pi^2/\epsilon}\,]}\right)^{1/2}\,.  
\end{equation}
Combined with Eq.\ \eqref{eq:147} this gives the transition probability
\begin{equation}
  \label{eq:304}
 p[(m,\dl) \leftrightarrow z]=  |(e_{m,\delta},\hat{u}_z^{(\epsilon,\delta)})|^2 = 
 \left(\frac{\epsilon}{\pi}\right)^{1/2}\,\frac{
e^{\ds-[\tilde{l}-\epsilon(m+\delta)]^2/\epsilon}}{\vt_3[\pi(\tl-\epsilon\dl)/\epsilon,\,
q=e^{\ds-\pi^2/\epsilon}\,]}\,.
\end{equation}
 The numerator of Eq.\ \eqref{eq:304} has its maximum for $\tl=\epsilon(m+\dl)$ with
 $\vt_3(\pi m,q)=\vt_3(0, q)$ for the denominator.

Here and below the use of $\vt_3$ with $q=e^{\ds -\pi^2/\epsilon}$ instead of
 $q=e^{\ds -\epsilon}$ has
 the following considerable advantage: Numerically one has
\begin{equation}
  \label{eq:152}
 q= e^{\ds -\pi^2} \approx 5.2\cdot 10^{\ds-5}\,.
\end{equation}
On the other hand, expanding $\vt_3(\zeta,q)$ in powers of $q$ (see Eq.\ \eqref{eq:141}) gives :
\begin{equation}
  \label{eq:153}
  \vt_3(\zeta,q)=1+2q\,\cos(2\zeta) + O(q^4)\,,
\end{equation}
so that $\vt_3$ in \eqref{eq:151} differs only very slightly from $1$! For $\epsilon \to 0$
we even have $q=e^{\ds -\pi^2/\epsilon} \to 0$!
We shall use this argument frequently in what follows. In that way one gets very
reasonable approximations for a number of expressions which contain $\vt$-functions.
This was previously pointed out by Kowalski, Rembieli\'{n}ski and Papaloucas \cite{kow2}.
 
The function \eqref{eq:140} may be written as
\begin{equation}
  \label{eq:160} u_z^{(\epsilon,\dl)}(\vp)=
  (\epsilon\pi)^{-1/4}\, e^{\ds- [(\vp-\theta)^2 +i\tl(\theta-2\vp)]/(2\epsilon)}
 \,
\vt_3[i\pi(\vp-z+i\epsilon\delta)/\epsilon ,\,e^{\ds-2\pi^2/\epsilon}\,]\,. 
\end{equation}
It yields the probability density
\begin{equation}
  \label{eq:149}
  p_z^{(\epsilon,\dl)}(\vp) = \frac{2\pi}{\sqrt{\epsilon\pi}}\,
e^{\ds-(\vp-\theta)^2/\epsilon}\,\frac{
|\vt_3[i\pi(\vp-z+i\epsilon\delta)/\epsilon,\,e^{\ds-2\pi^2/\epsilon}\,]|^2}{
\vt_3[\pi(\tl-\epsilon\dl)/\epsilon,\,e^{\ds -\pi^2/\epsilon}\,]}\,.
\end{equation}
Notice that 
\begin{equation}
  \label{eq:161}
  (\epsilon\pi)^{-1/2}\,
e^{\ds-(\vp-\theta)^2/\epsilon} \to \delta(\vp-\theta) \mbox{ for } \epsilon \to 0\,\,,
\end{equation}
so that in the classical limit we have $\vp\to \theta$ and the argument of $\vt_3$
in the numerator of Eq.\ \eqref{eq:149} approaches $\pi(\tl-\epsilon\delta)/\epsilon$\,.

For the scalar product of two coherent states $u_z^{(\epsilon,\dl)}(\vp)$ 
 we get
\begin{eqnarray}
  \label{eq:154}
&&  (u_{z_1}^{(\epsilon,\delta)},u_{z_2}^{(\epsilon,\delta)})\\ && = \frac{1}{2\pi}\left(
\frac{\epsilon}{\pi}\right)^{1/2}\,e^{\ds[-(\tl_1 -\epsilon\dl)^2/(2\epsilon)
-(\tl_2-\epsilon\dl)^2/(2\epsilon)]}\,e^{\ds i[(\theta_1-\theta_2)\dl
 -(\theta_1\tl_1-\theta_2\tl_2)/(2\epsilon)]}\times
\nonumber \\&& \times \vt_3[(z_1^*-z_2+2i\epsilon\dl)/2,\,e^{\ds-\epsilon}\,]\nonumber 
\end{eqnarray}
which reduces to the expression \eqref{eq:150}   for $z_2=z_1$\,.
\subsection{Coherent wave functions holomorphic in $z$}
 In the case of the conventionial coherent states \eqref{eq:139} or \eqref{eq:155} it can
have advantages to deal with wave functions which are holomorphic in the variable $z$ and
incorporate the non-holomorphic factor $e^{\ds -|z|^2/(2\epsilon)}$ into the measure of the
integral \eqref{eq:157}. In that way one  obtains Bargmann-Segal Hilbert spaces
of holomorphic functions \cite{barg}.

 Similarly one can split off a non-analytic factor from the 
function \eqref{eq:143} and define 
\begin{equation}
  \label{eq:162}
 w_z^{(\epsilon,\delta)}(\vp) = e^{\ds i\vp\delta}\,\vt_3[(\vp-z+i\epsilon\delta)/2,\,e^{\ds
 -\epsilon/2}\,]\,, 
\end{equation}
as the in $z$ holomorphic part, which, according to Eq.\ \eqref{eq:142}, can also be written as
\begin{align}
  \label{eq:163}
&  w_z^{(\epsilon,\delta)}(\vp)= \left(\frac{2\pi}{\epsilon}\right)^{1/2}\,
e^{\ds i\vp\delta}\,
e^{\ds -(\vp-z+i\epsilon\delta)^2/(2\epsilon)}\,\vt_3[i\pi(\vp-z+i\epsilon\delta)/\epsilon,\,
e^{\ds
 -2\pi^2/\epsilon}\,] \\  &= \left(\frac{2\pi}{\epsilon}\right)^{1/2}\,
e^{\ds-[(\vp-\theta)^2-(\tl-\epsilon\dl)^2]/(2\epsilon)}\,e^{\ds i[\tl(\vp-\theta)/\epsilon
+\theta\dl]}\,\vt_3[i\pi(\vp-z+i\epsilon\delta)/\epsilon,\,
e^{\ds
 -2\pi^2/\epsilon}\,]\,.\nonumber\end{align}

Similar to the $H_n(\xi)$--generating series \eqref{eq:324} the function \eqref{eq:162} may be
interpreted as a generating one for the basis $e_{n,\dl}(\vp)$ (with $\epsilon =1$):
\begin{equation}
  \label{eq:325}
   w_z^{(\epsilon=1,\delta)}(\vp)= \sum_{n \in \mathbb{Z}}e^{\ds -n^2/2}\,[\eta(z)\,
e^{\ds -\dl}]^{\ds n}\,e_{n,\dl}(\vp)\,,~~\eta(z)=e^{\ds-iz}\,.
\end{equation}

We have the scalar product
\begin{eqnarray}
  \label{eq:164}
  ( w_z^{(\epsilon,\delta)}, w_z^{(\epsilon,\delta)})&=& \vt_3[i(\tl-\epsilon\dl),\,
e^{\ds-\epsilon}\,] \\
 &=&\left(\frac{\pi}{\epsilon}\right)^{1/2}\,e^{\ds (\tl-\epsilon\dl)^2/\epsilon}\,
 \vt_3[\pi(\tl-\epsilon\dl)/\epsilon,\,
e^{\ds-\pi^2/\epsilon}\,]\label{eq:165}\\ &\equiv& (N_z^{(\epsilon,\dl)})^{\ds-2}\,,~
N_z^{(\epsilon,\dl)} >0\,, \nonumber
\end{eqnarray}
and therefore the normalized wave functions 
\begin{equation}
  \label{eq:166}
  \hat{w}_z^{(\epsilon,\dl)}= N_z^{(\epsilon,\dl)}\,w_z^{(\epsilon,\dl)}(\vp)\,.
\end{equation}
The scalar product between two different states is given by
\begin{eqnarray}
  \label{eq:167}
  (w_{z_1}^{(\epsilon,\dl)},w_{z_2}^{(\epsilon,\dl)}) &=& \vt_3[(z_1^*-z_2 +2i\epsilon\dl)/2,
\, e^{\ds-\epsilon}\,] \\
&=& \left(\frac{\pi}{\epsilon}\right)^{1/2}e^{\ds -(z_2^*-z_1+2i\epsilon\dl)^2/(4\epsilon)}\,
\vt_3[i\pi(z_1^*-z_2 +2i\epsilon\dl)/(2\epsilon),\,e^{\ds-\pi^2/\epsilon}\,]. \nonumber
\end{eqnarray}
As
\begin{equation}
  \label{eq:169}
  (e_{m,\dl},w_z^{(\epsilon,\dl)})=f_{m,\dl}(z)=e^{\ds-\epsilon\,( m^2/2 +m\dl)}
e^{\ds-imz}=
 e^{\ds-\epsilon\, m^2/2 +m(\tl-\epsilon\dl)}\,
e^{\ds-im\theta}\,,
\end{equation}
we have the completeness relation (see Eq.\ \eqref{eq:158})
\begin{equation}
  \label{eq:170}
\int_{\mathbb{R}}\frac{d\tl}{\sqrt{\epsilon\pi}}\,e^{\ds -(\tl-\epsilon\dl)^2/\epsilon}
\int_{S^1}\frac{d\theta}{2\pi}\, (e_{m_1,\dl},w_z^{(\epsilon,\dl)})(w_z^{(\epsilon,\dl)},
 e_{m_2,\dl})= \dl_{m_1m_2}\,.
\end{equation}
The functions $f_{m,\dl}(z)$ therefore form an orthornomal basis of a Hilbert
 space $\mathcal{H}_{\vt}$ of functions $\tilde{f}(z)$ 
holomorphic in the strip $z \in S^1+i\mathbb{R}$  with the scalar product
\begin{equation}
  \label{eq:273}
  (\tilde{f}_1,\tilde{f}_2)_z \equiv \int_{\mathbb{R}}\frac{d\tl}{\sqrt{\epsilon\pi}}\,e^{\ds
 -(\tl-\epsilon\dl)^2/\epsilon}
\int_{S^1}\frac{d\theta}{2\pi}\,\tilde{f}_1^*(z)\,,\tilde{f}_2(z)\,,
\end{equation}
so that the functions $\tilde{f}(z) \in \mathcal{H}_{\vt}$ may be expanded as
\begin{equation}
  \label{eq:274}
  \tilde{f}(z)=\sum_{n\in\mathbb{Z}}\tilde{c}_n\,f_{n,\dl}(z)\,,~~\tilde{c}_n=(f_{n,\dl},
\tilde{f})_z\,.
\end{equation}

The relation $\eqref{eq:169}$ provides a unitary mapping between the Hilbert space of functions
$f(\vp)$ with the scalar product \eqref{eq:22} and the Hilbert space $\mathcal{H}_{\vt}$:\\
If
\begin{equation}
  \label{eq:275}
  f(\vp)=\sum_{n\in \mathbb{Z}} b_n\,e_{n,\dl}(\vp)\,,~~b_n=(e_{n,\dl},f)\,,
\end{equation}
it follows that
\begin{equation}
  \label{eq:276}
  (f,w_z^{(\epsilon,\dl)})=\sum_{n\in \mathbb{Z}} b_n^*\,f_{n,\dl}(z)=\tilde{f}(z)\,,
\end{equation}
from which one infers that
\begin{equation}
  \label{eq:277}
  b_n^*=(f_{n,\dl},\tilde{f})_z\,.
\end{equation}
Unitarity can be seen from
\begin{equation}
  \label{eq:278}
  (f,f)=\sum_{n\in\mathbb{Z}}\,|b|^2 =(\tilde{f},\tilde{f})_z\,.
\end{equation}
and the inverse mapping:

If one has
\begin{equation}
  \label{eq:279}
  \tilde{f}(z)=\sum_{n\in\mathbb{Z}}\tilde{c}_n\,f_{n,\dl}(z)\,,~~\tilde{c}_n =(f_{n,\dl},
\tilde{f})_z\,,
\end{equation}
then the inverse mapping is
\begin{equation}
  \label{eq:280}
  f(\vp)=\sum_{n\in\mathbb{Z}}\tilde{c}_n^*\,e_{n,\dl}(\vp)=\int_{\mathbb{R}}\frac{d\tl}{
\sqrt{\epsilon
\pi}}\,e^{\ds -(\tl-\epsilon\dl)^2/\epsilon}\int_{S^1}\frac{d\theta}{2\pi}\,\tilde{f}^*(z)
\sum_{n\in\mathbb{Z}}f_{n,\dl}(z)e_{n,\dl}(\vp)=(\tilde{f},w_z^{(\epsilon,\dl)}(\vp))_z\,.
\end{equation}

If we replace the unnormalized wave functions \eqref{eq:162} by the normalized ones
\eqref{eq:166}, then the completeness relation \eqref{eq:170} takes the respective forms
\begin{align}& 
\int_{\mathbb{R}}\frac{d\tl}{\sqrt{\epsilon\pi}}\,e^{\ds -(\tl-\epsilon\dl)^2/\epsilon}\,
\vt_3[i(\tl-\epsilon\dl),e^{\ds-\epsilon}\,]\int_{S^1}\frac{d\theta}{2\pi}\,
 (e_{m_1,\dl},\hat{w}_z^{(\epsilon,\dl)})(\hat{w}_z^{(\epsilon,\dl)},
 e_{m_2,\dl})\label{eq:292} \\ &=
  \int_{\mathbb{R}}\frac{d\tl}{\epsilon} \vt_3[\pi(\tl-\epsilon\dl)/\epsilon,e^{\ds-
\pi^2/\epsilon}\,]\,
\int_{S^1}\frac{d\theta}{2\pi}\, (e_{m_1,\dl},\hat{w}_z^{(\epsilon,\dl)})(\hat{w}_z^{(
\epsilon,\dl)},
 e_{m_2,\dl})= \dl_{m_1m_2}\,. \nonumber
\end{align}

Finally, the reproducing kernel $K(z_1,z_2)$
is given by
\begin{equation}
  \label{eq:168}
  K(z_1,z_2)=\sum_{n \in \mathbb{Z}} f_{n,\dl}^*(z_1)f_{n,\dl}(z_2) = \vt_3[(z_1^*-z_2+
2i\epsilon\dl)/2,
e^{\ds-\epsilon}\,] = (w_{z_1}^{(\epsilon,\dl)},w_{z_2}^{(\epsilon,\dl)})\,.
\end{equation}
It fulfills the usual properties:
\begin{eqnarray}
  \label{eq:291}
K(z_2,z_1)&=& K^*(z_1,z_2)\,, \\
  \int_{S^1\times \mathbb{R}}d\mu(\theta,\tl)^{(\epsilon,\dl)}\, K(z_1,z)K(z,z_2) &=&
 K(z_1,z_2)\,,
\nonumber \\ \int_{S^1\times \mathbb{R}}d\mu(\theta,\tl)^{(\epsilon,\dl)}\,K(z,z_1)\,
f_{m,\dl}(z) &=& f_{m,\dl}(z_1)\,, \nonumber \\
d\mu(\theta,\tl)^{(\epsilon,\dl)}&=& \frac{e^{\ds -(\tl-\epsilon\dl)^2/\epsilon}\,d\tl}{
\sqrt{\epsilon
\pi}}\,\frac{d\theta}{2\pi}\,. \nonumber
\end{eqnarray}
\subsection{Expectation values}
Next we have to calculate the expectation values and the mean square fluctuations of
the observables $C=\cos\vp,\,S=\sin\vp$ and $L=\hbar\tilde{L}=(\hbar/i)\partial_{\vp}$
with respect to the normalized wave functions \eqref{eq:159} or \eqref{eq:166}.
As the $\dl$-dependence will always be in the wave function, the index $\dl$ of the operator
$\tilde{L}_{\dl}$ will be dropped. Several of the following expectation values have been
discussed by Kowalski et al.\ \cite{kow2} for the special cases $\dl=0$ , $\dl=1/2$ and
$\epsilon =1$. These authors require $T$-invariance (see the discussion in the final part
of Sec.\ 1 above).

It is convenient to start with the form \eqref{eq:162} of the wave function, with its
 normalization factor
\eqref{eq:164} and use the identity \eqref{eq:142} later:

Defining
\begin{equation}
  \label{eq:173}
  U= e^{\ds-i\vp}\,,~~~U^{\dagger} = e^{\ds i\vp}\,,
\end{equation}
we get
\begin{equation}
  \label{eq:174}
  \langle U \rangle_z^{(\epsilon,\dl)} \equiv (\hat{w}_z^{(\epsilon,\dl)},U\hat{w}_z^{(
\epsilon,\dl)}) = e^{\ds-i\theta}\,e^{\ds-\epsilon/4}\,\frac{\vt_2[i(\tl-\epsilon\dl),\,e^{\ds
 -\epsilon}\,]}{\vt_3[i(\tl-\epsilon\dl),\,e^{\ds
 -\epsilon}\,]}\,,
\end{equation}
where
\begin{equation}
  \label{eq:175}
  \vt_2(\zeta,q=e^{\ds i\pi\tau})\equiv \vt_2(\zeta|\tau) = \sum_{n \in \mathbb{Z}}
q^{\ds (n+1/2)^2}\,e^{\ds i(2n+1)\zeta}\,,~~\Im(\tau) >0\,,~\vt_2(-\zeta,q) = \vt_2(\zeta,q)\,.
\end{equation}
Instead of the identity \eqref{eq:142} we now have the following one:
\begin{equation}
  \label{eq:176}
  \vt_2(\zeta|\tau) =  (-i\tau)^{-1/2}e^{\ds \zeta^2/(i\pi\tau)}\,\vt_4(
\zeta/\tau|-1/\tau)\,,
\end{equation}
with
\begin{equation}
  \label{eq:177}
  \vt_4(\zeta,q) = \sum_{n \in \mathbb{Z}}(-1)^n q^{\ds n^2}e^{\ds 2ni\zeta}= 1+
2\sum_{n=1}^{\infty}(-1)^n \cos 2n\zeta\,. 
\end{equation}
Thus, Eq.\ \eqref{eq:174} may also be written as
\begin{equation}
  \label{eq:172}
  \langle U \rangle_z^{(\epsilon,\dl)}  = e^{\ds-i\theta}\,e^{\ds-\epsilon/4}\,
\frac{\vt_4[\pi(\tl-\epsilon\dl)/\epsilon,\,e^{\ds
 -\pi^2/\epsilon}\,]}{\vt_3[\pi(\tl-\epsilon\dl)/\epsilon,\,e^{\ds
 -\pi^2/\epsilon}\,]}\,.
\end{equation}
In the same way we get
\begin{equation}
  \label{eq:178}
 \langle U^{\dagger} \rangle_z^{(\epsilon,\dl)}  = e^{\ds i\theta}\,e^{\ds-\epsilon/4}\,
\frac{\vt_4[\pi(\tl-\epsilon\dl)/\epsilon,\,e^{\ds
 -\pi^2/\epsilon}\,]}{\vt_3[\pi(\tl-\epsilon\dl)/\epsilon,\,e^{\ds
 -\pi^2/\epsilon}\,]}\,,  
\end{equation}
so that
\begin{eqnarray}
  \label{eq:179}
  \langle C \rangle_z^{(\epsilon,\dl)} & =& \cos \theta\,e^{\ds-\epsilon/4}\,
\frac{\vt_4[\pi(\tl-\epsilon\dl)/\epsilon,\,e^{\ds
 -\pi^2/\epsilon}\,]}{\vt_3[\pi(\tl-\epsilon\dl)/\epsilon,\,e^{\ds
 -\pi^2/\epsilon}\,]}\,, \\ \label{eq:180}
 \langle S \rangle_z^{(\epsilon,\dl)} & =& \sin \theta\,e^{\ds-\epsilon/4}\,
\frac{\vt_4[\pi(\tl-\epsilon\dl)/\epsilon,\,e^{\ds
 -\pi^2/\epsilon}\,]}{\vt_3[\pi(\tl-\epsilon\dl)/\epsilon,\,e^{\ds
 -\pi^2/\epsilon}\,]}\,.
\end{eqnarray}
Combining
\begin{equation}
  \label{eq:181}
  \vt_4(\zeta,q) = 1-2q\,\cos (2\zeta) + O(q^4)
\end{equation}
with the Eqs.\ \eqref{eq:153} and \eqref{eq:152} we have the very good approximation
\begin{equation}
  \label{eq:182}
  \frac{\vt_4[\pi(\tl-\epsilon\dl)/\epsilon,\,e^{\ds
 -\pi^2/\epsilon}\,]}{\vt_3[\pi(\tl-\epsilon\dl)/\epsilon,\,e^{\ds
 -\pi^2/\epsilon}\,]} = 1 -4\,e^{\ds-\pi^2/\epsilon}\cos[2\pi(\tl-\epsilon\dl)/\epsilon]
+O(e^{\ds-2\pi^2/\epsilon})\,.
\end{equation}
Notice that $O(e^{\ds -2\pi^2}) = 10^{-9}$\,!

For the expectation value of the orbital angular momentum operator 
we get
\begin{equation}
  \label{eq:183}
  \langle \tilde{L} \rangle_z^{(\epsilon,\dl)} = \dl + \frac{1}{2}\,\frac{d\vt_3[i(\tl-\epsilon
\dl),\,e^{\ds-\epsilon}\,]/d\tl}{\vt_3[i(\tl-\epsilon
\dl),\,e^{\ds-\epsilon}\,]}\,.
\end{equation}
Again using the identity \eqref{eq:142} yields
\begin{equation}
  \label{eq:184}
  \epsilon\, \langle \tilde{L} \rangle_z^{(\epsilon,\dl)}= \tl + \epsilon \dl +\frac{\pi}{2}
\frac{\vt_3^{\prime}[\pi(\tl-\epsilon\dl)/\epsilon,\,e^{\ds-\pi^2/\epsilon}\,]}{\vt_3[\pi(
\tl-\epsilon\dl)/
\epsilon,\,e^{\ds-\pi^2/\epsilon}\,]}\,,
\end{equation}
where $\vt_3^{\prime}(\zeta,q)$ means the derivative with respect to the full argument $\zeta$.
As
\begin{equation}
  \label{eq:185}
  \frac{\vt_3^{\prime}[\pi(\tl-\epsilon\dl)/\epsilon,\,e^{\ds-\pi^2/\epsilon}\,]}{\vt_3[
\pi(\tl-\epsilon\dl)/
\epsilon,\,e^{\ds-\pi^2/\epsilon}\,]} = -4\,e^{\ds-\pi^2/\epsilon}\,\sin[2\pi(\tl-\epsilon\dl)
/\epsilon] + O(e^{\ds -2\pi^2})\,,
\end{equation}
the last term in Eq.\ \eqref{eq:184} constitutes only a very small correction which vanishes
in the classical limit $\epsilon \to 0$. 

It follows from Eq.\ \eqref{eq:184} that in classical limits of matrix elements
 now the product $\epsilon\,\tilde{L}$ should be kept fixed! See the corresponding remarks
for $\tilde{P}$ after Eq.\ \eqref{eq:323}.

From
\begin{equation}
  \label{eq:186}
  \langle U^2 \rangle_z^{(\epsilon,\dl)} = e^{\ds-\epsilon}\,e^{\ds-2i\theta}\,,~~~
 \langle (U^{\dagger})^2 \rangle_z^{(\epsilon,\dl)} = e^{\ds-\epsilon}\,e^{\ds 2i\theta}\,,
\end{equation}
we get
\begin{eqnarray}
  \label{eq:187}
  \langle C^2 \rangle_z^{(\epsilon,\dl)} &=&\frac{1}{2} +e^{\ds-\epsilon}\,(\cos^2 \theta -
\frac{1}{2})\,, \\ ((\Delta C)_z^{(\epsilon,\dl)})^2 &=& \frac{1}{2} +e^{\ds-\epsilon}\,(\cos^2
 \theta -\frac{1}{2})-\label{eq:188}\\ &&-e^{\ds-\epsilon/2}\,\cos^2 \theta\,
\left(\frac{\ds \vt_4[\pi(\tl-\epsilon\dl)/\epsilon,\,e^{\ds
 -\pi^2/\epsilon}\,]}{\ds \vt_3[\pi(\tl-\epsilon\dl)/\epsilon,\,e^{\ds
 -\pi^2/\epsilon}\,]}\right)^2\,. \nonumber \\
 \langle S^2 \rangle_z^{(\epsilon,\dl)} &=&\frac{1}{2} +e^{\ds-\epsilon}\,(\sin^2 \theta -
\frac{1}{2})\,,\label{eq:189} \\ 
 ((\Delta S)_z^{(\epsilon,\dl)})^2 &=& \frac{1}{2} +e^{\ds-\epsilon}\,(\sin^2
 \theta -\frac{1}{2})-\label{eq:190}\\ &&-e^{\ds-\epsilon/2}\,\sin^2 \theta\,
\left(\frac{\ds \vt_4[\pi(\tl-\epsilon\dl)/\epsilon,\,e^{\ds
 -\pi^2/\epsilon}\,]}{\ds \vt_3[\pi(\tl-\epsilon\dl)/\epsilon,\,e^{\ds
 -\pi^2/\epsilon}\,]}\right)^2\,. \nonumber
\end{eqnarray}
It follows that
\begin{equation}
  \label{eq:191}
  ((\Delta C)_z^{(\epsilon,\dl)})^2+ ((\Delta S)_z^{(\epsilon,\dl)})^2 = 1-
e^{\ds-\epsilon/2}\,
\left(\frac{\ds \vt_4[\pi(\tl-\epsilon\dl)/\epsilon,\,e^{\ds
 -\pi^2/\epsilon}\,]}{\ds \vt_3[\pi(\tl-\epsilon\dl)/\epsilon,\,e^{\ds
 -\pi^2/\epsilon}\,]}\right)^2\,.
\end{equation}

In view of  the relation \eqref{eq:182} it makes good sense to approximate that ratio $\vt_4/
\vt_3$ by
$1\,$! In that approximation the fluctuations \eqref{eq:188} have their maximum for $\cos \theta
=0$, namly $(1-e^{\ds -\epsilon})/2$. 

We further have the following relations
\begin{eqnarray}
  \label{eq:192}
  \epsilon^2\,\langle \tilde{L}^2 \rangle_z^{(\epsilon,\dl)} &=&\frac{\epsilon}{2} +(\tl+\epsilon
\dl)^2+\pi(\tl+\epsilon\dl)\frac{\vt_3^{\prime}}{\vt_3} +\frac{\pi^2}{4}\frac{\vt_3^{\prime
 \prime}}{\vt_3}\,,\\ \epsilon^2\,((\Delta \tilde{L})_z^{(\epsilon,\dl)})^2 &=&
 \frac{\epsilon}{2}+\frac{\pi^2}{4}\left(\frac{\vt_3^{\prime \prime}}{\vt_3}-
\frac{\vt_3^{\prime\,
2}}{\vt_3^2}\right)\,,\label{eq:193}
\end{eqnarray} where the function $\vt_3$ and its derivatives mean the same as in Eq.\
\eqref{eq:184}.

With the approximation \eqref{eq:153} the relation \eqref{eq:193} takes the form
\begin{equation}
  \label{eq:194}
  \epsilon^2\,((\Delta \tilde{L})_z^{(\epsilon,\dl)})^2 = \frac{\epsilon}{2} -2\pi^2\,e^{\ds
-\pi^2/\epsilon}\,\cos[2\pi(\tl-\epsilon\dl)/\epsilon] +O(e^{\ds-2\pi^2/\epsilon})\,.
\end{equation}

Finally we get for the correlation function \eqref{eq:64} with $A= C,\,B=\tilde{L}$ and
$\psi =\hat{w}_z^{(\epsilon,\dl)}\,$:
\begin{eqnarray}
  \label{eq:195}
  \epsilon\,\langle S_z^{(\epsilon,\dl)}(C,\tilde{L}) \rangle_z^{(\epsilon,\dl)}&=&
\frac{\pi}{2}\,e^{\ds -\epsilon/4}\,\cos\theta \,\frac{\vt_4}{\vt_3}\left(
\frac{\vt_4^{\prime}}{\vt_4}-\frac{\vt_3^{\prime}}{\vt_3}\right)\\
& =&4\pi\,e^{\ds-\pi^2/\epsilon}\,
e^{\ds-\epsilon/4}\,\cos \theta\,\sin[2\pi(\tl-\epsilon\dl)/\epsilon)]+O(e^{\ds-2\pi^2/\epsilon})
\,.\nonumber
\end{eqnarray}

The states \eqref{eq:162} are no minimal uncertainty states for the operators $C,\,S$ and
$L$. This can already be seen in the
$q^0$ approximation mentioned above: For the left-hand side of the relation \eqref{eq:63}
we here get with $\epsilon =1$:
\begin{equation}
  \label{eq:196}
((\Delta C)_z^{(\epsilon,\dl)})^2\, ((\Delta \tilde{L})_z^{(\epsilon,\dl)})^2 = \frac{1}{2}\,[
\frac{1}{2}+e^{\ds-1}(\cos^2 \theta -1/2)-e^{\ds-1/2}\cos^2 \theta]\,,
\end{equation}
whereas the right-hand side is (the square of the correlation \eqref{eq:195} is negligible here)
\begin{equation}
 \label{eq:215}
  \frac{1}{4}e^{\ds-1/2}\sin^2 \theta.
\end{equation}
The difference between the two sides has its minimum for $\cos \theta=0$. In that case the
 inequality reads
\begin{equation}
  \label{eq:197}
  (1-e^{\ds-1})/4 > e^{\ds-1/2}/4\,.
\end{equation}

We shall see below for which self-adjoint operators the coherent states \eqref{eq:162}
are minimal uncertainty states.

For physical applications the dimensionless operator $\tilde{L}$ should be multiplied by
$\hbar$ in the formulae above and $\tl + \dl$ replaced by $(l+\mu)/\hbar$:
\begin{equation}
  \label{eq:198}
  \tilde{L} =L/\hbar\,\,,~~~\tl\pm \dl = (l\pm \mu)/\hbar\,,~\mu \equiv \hbar\,\dl\,.
\end{equation}

Notice that the quasi orbital momentum $\dl$, a genuine quantum quantity, appears in
 the formulae above generally in the
form $\epsilon \delta$, i.e.\ it vanishes in the classical limit as it should.

As to the mathematics: The ratios of $\vt$-functions appearing above may be expressed by
Jacobi's elliptic functions. Examples will be given in Appendix C.
\section{Holomorphic coherent states on the circle generated as eigenstates 
of composite  annihilation operators}
Like the conventional Schr\"{o}dinger-Glauber coherent states the states \eqref{eq:162}
 may also be generated as eigenstates of certain ``annihilation''
or ``ladder'' operators.
This was first done ``by hand'' by Kowalski et al.\ \cite{kow2} for the special cases
 $\dl =0$ and $\dl=1/2$, because the authors imposed $T$-invariance.

Following the work of Hall on coherent states \cite{hall1} Thiemann and coworkers \cite{thie1,
thie2} systematically constructed coherent states for the groups $U(1)=SO(2) =S^1$ (with
 $\dl=0$) and $SU(2)$
in connection with  problems of the classical limit for Loop Quantum Gravity.
Finally, Hall and Mitchell \cite{hall2} discussed the case of general $S^n$. (The method of Ch.\
4 generalizes to $n$-dimensional tori $T^n$ for $n>1$. Only for $n=1$ the two methods give the
same results because $T^1=S^1$.)
The basic idea of Hall was, to extend certain functions of the configuration variable $\tilde{q}$
to holomorphic functions of the complex variable $z=\tilde{q}+i\tilde{p}$ (I here consider only
1-dimensional configuration spaces). Let us see how this works for the conventional coherent
states \eqref{eq:155}:

First let us change the normalization of the functions \eqref{eq:155} without changing their
essential properties. The slightly modified functions
\begin{equation}
  \label{eq:199}
  v_z^{(\epsilon)}(\xi) = \frac{e^{\ds-\tilde{p}^2/(2\epsilon)}}{(\epsilon\pi)^{1/4}}\,
e^{\ds-(\xi-z)^2/(2\epsilon)}
\end{equation}
have the properties 
\begin{eqnarray}
  \label{eq:200}
  \int_{\mathbb{R}}d\xi\,{v_z^{(\epsilon)}}^*(\xi)\,v_z^{(\epsilon)}(\xi)& =&1\,, \\
\int_{\mathbb{R}^2}\frac{d\tilde{q}\,d\tilde{p}}{2\pi\epsilon}
{v_z^{(\epsilon)}}^*(\xi_1)\,v_z^{(\epsilon)}(\xi_2)&=& \dl(\xi_1-\xi_2)\,. \label{eq:201}
\end{eqnarray}
The functions \eqref{eq:199} can be generated in the following way:

The $\dl$-function
\begin{equation}
  \label{eq:202}
  \dl_{\tilde{q}}(\xi) =\dl(\xi-\tilde{q})=\frac{1}{2\pi}\int_{-\infty}^{+\infty} d\tilde{p}
\,e^{\ds i\tilde{p}(\xi-\tilde{q})}
\end{equation}
may formally be considered as an ``eigenfunction'' of the position operator $\tilde{Q} =\xi$
with\\ ``eigenvalue'' $\tilde{q}$:
\begin{equation}
  \label{eq:203}
  \tilde{Q}\,\dl_{\tilde{q}}(\xi) = \xi\,\dl_{\tilde{q}}(\xi) =\tilde{q}\,\dl_{\tilde{q}}(\xi)\,.
\end{equation}
Applying the operator
\begin{equation}
  \label{eq:204}
  C_{P}=e^{\ds-\epsilon \tilde{P}^2/2}\,,~~\tilde{P}=\frac{1}{i}\partial_{\xi}\,,
\end{equation}
to the generalized function \eqref{eq:202} yields
\begin{equation}
  \label{eq:205}
  C_P\,\dl_{\tilde{q}}(\xi) = \frac{1}{2\pi}\,\int_{-\infty}^{+\infty} d\tilde{p}\,
e^{\ds-\epsilon\,\tilde{p}^2/2}
\,e^{\ds i\tilde{p}(\xi-\tilde{q})}=\frac{1}{\sqrt{2\epsilon\pi}}\,e^{\ds-(\xi-\tilde{q}
)^2/(2\epsilon)}\,,
\end{equation}
which is - up to a constant - just the limit $\Im (z) \to 0$ of the holomorphic factor
\begin{equation}
  \label{eq:206}
e^{\ds-(\xi-z)^2/(2\epsilon)}  
\end{equation}
of the coherent state \eqref{eq:199}. Conversely, we only  have to replace $\tilde{q}$ in
Eq.\  \eqref{eq:205}
by $z$!

Using the relation
\begin{equation}
  \label{eq:207}
  e^{\ds A}\,B\,e^{\ds-A} = B+[A,B]+\frac{1}{2!}[A,[A,B]] + \cdots
\end{equation}
yields
\begin{equation}
  \label{eq:208}
  C_P\,\tilde{Q}\,C^{-1}_P =\tilde{Q}+i\epsilon \tilde{P}\,,
\end{equation}
which, according to Eq.\ \eqref{eq:234},  is  the annihilation operator with 
 eigenvalues $z$.
 
The method works for the coherent states \eqref{eq:162} in the following way:

Le $f(\vp)$ be a  smooth test function with the property
\begin{equation}
  \label{eq:209}
  f(\vp+2\pi)=e^{\ds i2\pi\dl}f(\vp)\,.
\end{equation}
The $\dl$-function for this type of test functions is
\begin{equation}
  \label{eq:210}
  \dl_{\theta}(\vp) = e^{\ds i(\vp-\theta)\dl}\sum_{n \in\mathbb{Z}}e^{\ds i n(\vp-\theta)}\,,
\end{equation}
because
\begin{equation}
  \label{eq:211}
  \int_{S^1}\frac{d\vp}{2\pi}\dl_{\theta}^*(\vp)\,f(\vp) =\sum_{n\in\mathbb{Z}}c_n e^{\ds i(n+
\dl)\theta} = f(\theta)\,,~~c_n= \int_{S^1}\frac{d\vp}{2\pi}e^{\ds-i(n+\dl)\vp}\,f(\vp)\,.
\end{equation}
As $\dl_{\theta}(\vp)$ is a complex functional here, one has to take its complex conjugate
in Eq.\ \eqref{eq:211} \cite{gel}.\\ (There should again be no confusion
 between the $\dl$-functional
\eqref{eq:210} and the parameter $\dl$ which characterizes the quasi-OAM!)

Applying the operator
\begin{equation}
  \label{eq:212}
  C_L =e^{\ds-\epsilon\tilde{L}^2/2}\,,~~\tilde{L}=\frac{1}{i}\partial_{\vp}\,,
\end{equation}
to the $\dl$-functional \eqref{eq:210} yields
\begin{eqnarray}
  \label{eq:213}
  C_L\dl_{\theta}(\vp) &=&\sum_{n \in \mathbb{Z}} e^{\ds-\epsilon(n+\dl)^2/2}\,e^{\ds i(n+\dl)(
\vp-\theta)}\\ &=& e^{\ds i(\vp-\theta)\dl-\epsilon\dl^2/2}\,\vt_3[(\vp-\theta+i\epsilon\dl),\,
e^{\ds-\epsilon/2}\,]\,. \nonumber
\end{eqnarray}
Replacing the real variable $\theta$ by the complex one $z=\theta+i\tl$ yields, up to a $\vp\,$-
independent factor, the holomorphic part of the functions \eqref{eq:143} or \eqref{eq:162},
respectively. Thus, the operator $C_L$ acts as a kind of ``complexifier'' \cite{thie1,hall2}.

That complexifying procedure has another intriguing aspect \cite{hall2}:

The configuration space $S^1$ may be parametrized by the two functions $\cos \theta$ and
$\sin \theta$ which obey
\begin{equation}
  \label{eq:221}
  \cos^2\theta +\sin^2\theta=1\,.
\end{equation}
If we replace $\theta$ by $z=\theta+i\tl$, we still have
\begin{equation}
  \label{eq:222}
  \cos^2z+\sin^2z=1\,,~~z=\theta+i\tl\,,~\theta \in \mathbb{R}\,\bmod{2\pi}\,,~\tl \in 
\mathbb{R}\,.
\end{equation}
Thus, we may characterize the phase space \eqref{eq:12} by the complex sphere $S^1_{\mathbb{C}}$
given by Eq.\ \eqref{eq:222}!

Using the commutation relations
\begin{equation}
  \label{eq:214}
  [\tilde{L},U] =-U\,\,,~~~~[\tilde{L},U^{\dagger}]=U^{\dagger}\,,~~U=e^{\ds \,-i\vp}\,,
\end{equation}
and the relation \eqref{eq:207} yields
\begin{equation}
  \label{eq:216}
  C_L\,U\,C^{-1}_L = e^{\ds \epsilon(\tilde{L}+1/2)}\,U =B_{\epsilon,\dl}\equiv B\,.
\end{equation}
It follows that
\begin{equation}
  \label{eq:217}
  B\,e_{n,\dl}(\vp) = e^{\ds \epsilon(n+\dl-1/2)}\,e_{n-1,\delta}(\vp)\,,~~B^{\dagger}\,
\,e_{n,\dl}(\vp) = e^{\ds \epsilon(n+\dl+1/2)}\,e_{n+1,\delta}(\vp)\,.
\end{equation}
That action of $B$ implies that the functions \eqref{eq:162} are eigenfunctions of
 the operator $B$:
\begin{equation}
  \label{eq:218}
  B\,w_z^{(\epsilon,\dl)}=\eta\,w_z^{(\epsilon,\dl)}\,,~~\eta =e^{\ds-iz}\,,~z=\theta+i\tl\,.
\end{equation}
This may be verified - using the relations \eqref{eq:217} - either by direct calculation
or from the ansatz
\begin{equation}
  \label{eq:219}
B\,\sum_{n \in \mathbb{Z}}c_n\, e_{n,\dl} =\eta\,\sum_{n \in \mathbb{Z}}c_n\,e_{n,\dl}\,,  
\end{equation}
which leads to the recursion formulae
\begin{eqnarray}
  \label{eq:220}
  c_{n+1}&=&\eta\,e^{\ds-\epsilon\,(n+\dl+1/2)}\,c_n\,, \\ c_n&=&\eta^n\,e^{\ds-\epsilon\, n^2/2
-n\,\epsilon
\dl}\,c_0 \nonumber \\ &=& e^{\ds-\epsilon\,n^2/2}\,e^{\ds 2ni(-z+i\epsilon\dl)/2}c_0\,.
\nonumber
\end{eqnarray}
Inserting these $c_n$ into Eq.\ \eqref{eq:219},  with $c_0=1$, yields $w_z^{(\epsilon,
\dl)}(\vp)$.

The operators $B$, $\,B^{\dagger}$ and $\tilde{L}$ have an interesting algebraic structure
 of their own \cite{kow2}:

It follows from the relations \eqref{eq:217} and \eqref{eq:214} that
\begin{eqnarray}
  \label{eq:223}
  B^{\dagger}\,B &=& e^{\ds -2\epsilon}\,B\,B^{\dagger}=e^{\ds-2\epsilon}\,
e^{\ds\epsilon(2\tilde{L}+1)}\,, \\
B^{\dagger}\,B\,e_{n,\dl}& =& e^{\ds \epsilon[2(n+\delta)-1]}\,e_{n,\dl}\,, \label{eq:239}\\
 {[} B, B^{\dagger} {]} &=& 2(\sinh\epsilon)\,e^{\ds 2\epsilon\tilde{L}}\, \label{eq:235} \\
{[}\tilde{L},B{]} &=& -B\,,~~ 
{[}\tilde{L},B^{\dagger}{]} = B^{\dagger} \,.\label{eq:238}
\end{eqnarray}
Recall that these operators act in a Hilbert space $L^2(S^1,d\vp/2\pi,\dl)$ the elements of
wich have the property \eqref{eq:42}.

If we define the self-adjoint operators
\begin{equation}
  \label{eq:224}
  K=B+B^{\dagger}\,,~~J=i(B^{\dagger}-B)\,,
\end{equation}
we have
\begin{equation}
  \label{eq:225}
  [K,J]=4i\,(\sinh\epsilon)\,e^{\ds 2\epsilon\tilde{L}}\,.
\end{equation}
Writing
\begin{equation}
  \label{eq:226}
  |\eta\rangle = \hat{w}_z^{(\epsilon,\dl)}
\end{equation}
for the normalized coherent states \eqref{eq:166}, we have the expectation values etc.:
\begin{eqnarray}
  \label{eq:227}
 \langle K\rangle_{\eta} \equiv \langle \eta|K|\eta\rangle &=& \eta + \eta^*
 = 2\cos\theta\,e^{\ds \tl}\,, \\
\langle J\rangle_{\eta} &=& i(\eta^* - \eta) = -2\sin\theta\,e^{\ds \tl}\,,\label{eq:228} \\
\tan\theta&=&-\frac{\langle J\rangle_{\eta}}{\langle K\rangle_{\eta}}\,,\label{eq:316} \\
\tl&=& \ln[(\langle K\rangle_{\eta}^2+  \langle J\rangle_{\eta}^2)/4]^{1/2}\label{eq:317} \\
(\Delta K)^2_{\eta}&=&(e^{\ds 2\epsilon}-1)\,\eta^*\eta=(e^{\ds 2\epsilon}-1)\,e^{\ds 2\tl}\,,
\label{eq:229} \\
(\Delta J)^2_{\eta}&=&(e^{\ds 2\epsilon}-1)\,\eta^*\eta=(e^{\ds 2\epsilon}-1)\,e^{\ds 2\tl}\,,
\label{eq:230}\\
\langle S_{\eta}(K,J)\rangle_{\eta}& =& 0\,,~~S_{\eta}(K,J)=(K\,J+J\,K)/2-\langle K\rangle_{\eta}
\langle J\rangle_{\eta} \label{eq:231} \, \\ \langle [K,J]\rangle_{\eta}&=&2i\,(e^{\ds
 2\epsilon}-1)\,
\eta^*\eta=2i\,(e^{\ds 2\epsilon}-1)\,e^{\ds 2\tl}\,.\label{eq:236} 
\end{eqnarray}
From the Eqs.\ \eqref{eq:229}-\eqref{eq:236} it follows that
\begin{equation}
  \label{eq:237}
(\Delta K)^2_{\eta}\,(\Delta J)^2_{\eta}=\frac{1}{4}\,|\langle [K,J]\rangle_{\eta}|^2\,.  
\end{equation}
This shows that the coherent states \eqref{eq:226} are minimal uncertainty states for the
self-adjoint operators \eqref{eq:224}!

Defining 
\begin{equation}
  \label{eq:240}
  A_{\epsilon,\dl} = (1+e^{\ds -2\epsilon})^{\ds -1/2}\,B_{\epsilon,\dl}\,,~~
 A_{\epsilon,\dl}^{\dagger} =
 (1+e^{\ds -2\epsilon})^{\ds -1/2}\,B_{\epsilon,\dl}^{\dagger}\,,
\end{equation}
\begin{equation}
  \label{eq:241}
  N_{\epsilon,\dl}=\tilde{L} +\frac{1}{2\epsilon}\ln(2\sinh \epsilon)\,,
\end{equation}
the commutation relation \eqref{eq:235} takes the form
\begin{equation}
  \label{eq:242}
  A_{\epsilon,\dl}\, A_{\epsilon,\dl}^{\dagger}-q\, A_{\epsilon,\dl}^{\dagger}\, A_{\epsilon,\dl}
 =q^{\ds -N_{\epsilon,\dl}}\,,~~q=e^{\ds - 2\epsilon}\,.
\end{equation}
This is one possible form of a so-called ``$q$-deformed oscillator algebra'' \cite{MacF,Bie,Buz,
Solo}. However, as
\begin{equation}
  \label{eq:243}
  (e_{n,\dl},N_{\epsilon,\dl}\,e_{n,\dl})=n+\dl +\frac{1}{2\epsilon}\ln(2\sinh \epsilon)\,,~~n\in
 \mathbb{Z}\,,
\end{equation}
the operator $N_{\epsilon,\dl}$ is neither bounded from below nor are its eigenvalues in
 general integers!
Numerically one has for $\epsilon =1$:
\begin{equation}
  \label{eq:244}
  \frac{1}{2}\ln(2\sinh(1)) = 0.427\,.
\end{equation}
By an appropriate choice of $\dl$ one can make the expectation value \eqref{eq:243} an integer.

Aspects of the representation theory of the algebra \eqref{eq:242} in the context of
 the euclidean
group $E(2)$ have been discussed by Woronowicz \cite{wor} and Rideau \cite{rid}. \newpage
\section{Time evolution of the coherent states}
Let us  have a brief look at the time evolution of the  states \eqref{eq:101} and 
\eqref{eq:162} under 
the action
of the Hamiltonian (see Sec.\ 1)
\begin{equation}
  \label{eq:245}
  H_{\dl}= \epsilon\frac{\hbar\omega}{2} \tilde{L_{\dl}}^2\,,~\tilde{L}_{\dl}
 =-i\partial_{\vp}\,,
\end{equation}
The corresponding unitary time evolution operator is
\begin{equation}
  \label{eq:246}
  U_{\dl}(t) = e^{\ds-i(H_{\dl}/\hbar)t}=e^{\ds-i\epsilon (\tilde{L_{\dl}}^2/2)\omega t}\,.
\end{equation}
 It is of general interest to determine the kernel (``propagator'') $K_t^{(\dl)}(\vp-\phi)$
 which solves the initial value problem
 \begin{equation}
   \label{eq:305}
   i\hbar\,\partial_t\psi(t,\vp)=H_{\dl}\psi(t,\vp)\,,~\psi(t=0,\vp)=\chi(\vp)\,,
 \end{equation}
namely
\begin{equation}
  \label{eq:306}
  \psi(t,\vp)=\int_{S^1}\frac{d\phi}{2\pi} K_t^{(\dl)}(\vp-\phi)\chi(\phi)\,,~\lim_{t
 \to 0} K_t^{(\dl)}(\vp-\phi)=\dl(\vp-\phi)\,,
\end{equation}
where $\dl(\vp-\phi)$ is given by Eq.\ \eqref{eq:210}.

Inserting the ansatz
\begin{equation}
  \label{eq:307}
  K_t^{(\dl)}(\vp) = \sum_{n\in \mathbb{Z}} k_n(t)\,e_{n,\dl}(\vp)
\end{equation}
into the Schr\"{o}dinger Eq.\ \eqref{eq:305} yields
\begin{equation}
  \label{eq:308}
  k_n(t) = e^{\ds -i\epsilon(n+\dl)^2\omega t/2}\,,
\end{equation}
so that
\begin{eqnarray}
  \label{eq:309}
  K_t^{(\dl)}(\vp-\phi)&=&\sum_{n\in \mathbb{Z}}e^{\ds -i\epsilon(n+\dl)^2\omega
 (t-i\eta)/2}\,e^{\ds i
(n+\dl)(\vp-\phi)}\\ &=&e^{\ds -i\epsilon \dl^2 \omega t/2}\,e^{\ds i(\vp-\phi)
\dl}\,\vt_3[(\vp-\phi-\epsilon
 \dl \omega
 t)/2,\, q=e^{\ds -i\epsilon \omega (t-i\eta)/2}]\,. \nonumber
\end{eqnarray}
In order to insure convergence of the series \eqref{eq:309} one has to give the time $t$
a small negative imaginary part $-i\eta, \eta > 0$, which is to be taken to 0 at the end of
the calculation \eqref{eq:306}. The situation is completely analogous to the case of the kernel
for a free particle in space where one has to proceed in the same way (i.e.\ $t\to t-i\eta$)
 when Fourier transforming \cite{reed2}. The kernel \eqref{eq:309} obviously has the required
property \eqref{eq:306} for $t-i\eta \to 0$.

Again using the identity \eqref{eq:142} one gets the following alternative form for the kernel
 \eqref{eq:309}
\begin{eqnarray}
  \label{eq:310}
   K_t^{(\dl)}(\vp-\phi)&=&\frac{\sqrt{2\pi}\,e^{\ds -i\epsilon \dl^2 \omega t/2
 }}{[i\epsilon \omega (t-i\eta)]^{1/2}}\, e^{\ds -(\vp-\phi-\epsilon
\dl \omega t)^2/(2i\epsilon \omega t)}\,e^{\ds i(\vp-\phi)\dl }\times \\
&& \times\vt_3\{\pi[\vp-\phi -\epsilon \dl \nonumber
\omega (t-i\eta)]/[\epsilon \omega (t-i\eta)],q=e^{\ds 2i\pi^2/[\epsilon \omega (t-i\eta)]}\}\,.
\end{eqnarray}
For $\epsilon=1$ and $\dl=0$ the expression simplifies considerably and takes a form similar
to  the kernel of a free particle (with mass $m=1/2$ and $\hbar =1$) in one space
 dimension \cite{reed2}:
\begin{equation}
  \label{eq:290}
  K_t(x-y)= (4\pi i t)^{-1/2}\,e^{\ds i(x-y)^2/(4t)}\,.
\end{equation}
 (Up to now I have assumed $t>0$. If $t<0$ the
formulae above change accordingly.)

For $\epsilon =1$ and $\dl=0$ the kernel \eqref{eq:310} is very closely related to the
corresponding kernel of the heat equation on the circle \cite{tay}.

If the function $\chi$ from Eq.\ \eqref{eq:305} has the expansion
\begin{equation}
  \label{eq:311}
  \chi(\phi)=\sum_{m\in\mathbb{Z}} c_m(\chi)\,e_{m,\dl}(\phi)\,,~~c_m(\chi) =(e_{m,\dl},\chi)\,,
\end{equation}
then the time evolution \eqref{eq:306} takes the form
\begin{equation}
  \label{eq:312}
  \psi(t,\vp) = e^{\ds -i\epsilon \dl^2 \omega t/2 }\,e^{\ds i\vp \dl}\,\sum_{m\in\mathbb{Z}}
e^{\ds -i\epsilon \omega (t-i\eta)m^2/2}\,e^{\ds im[\vp-\epsilon \dl \omega
 (t-i\eta)]}c_m(\chi)\,.
\end{equation}
In the case of the states \eqref{eq:101} the coefficients $c_m$ are given by Eq.\
 \eqref{eq:119}, so that we have
 \begin{eqnarray}
   \label{eq:313}
   \psi_{\alpha,n+\dl}(t,\vp)&=&e^{\ds -i\epsilon \dl^2 \omega t/2 }\,e^{\ds i(\vp-\alpha) \dl}
\times \\ && \times
\sum_{m\in\mathbb{Z}}e^{\ds -i\epsilon \omega (t-i\eta)m^2/2}\,e^{\ds im[\vp-\alpha-\epsilon
 \dl \omega (t-i\eta)]}\,J_{m-n}(\sigma)/\sqrt{I_0(2s)}\,. \nonumber
 \end{eqnarray}

The time evolution of the states \eqref{eq:162} can be obtained more directly:

Applying $U(t)$ from Eq. \eqref{eq:246} to them  yields
\begin{equation}
  \label{eq:289}
  U(t)w_z^{(\epsilon,\dl)}(\vp)=e^{\ds -i\epsilon\dl^2\,\omega t/2}\,e^{\ds i\vp\dl}\,
\vt_3[(\vp-z-\epsilon \dl\omega t+i\epsilon\dl)/2,q=e^{\ds-\epsilon(1+i\omega t)/2}]\,.
\end{equation}
Another possibility is to 
apply $U(t)$ to the state \eqref{eq:213} which amounts to the replacement
\begin{equation}
  \label{eq:247}
  \epsilon \to \epsilon (1+i\omega t)\,,
\end{equation}
because
\begin{equation}
  \label{eq:248}
  U(t)\,C_L=e^{\ds -\epsilon(1+i\omega t)\tilde{L}^2/2}\,.
\end{equation}
The result \eqref{eq:289} means that $U(t)$ generates a time-dependent phase for the
 function \eqref{eq:162},
 replaces the angle $\theta$ in the argument of $\vt_3$ by the time-dependent $\theta + 
(\epsilon\,\omega \dl)t$ one and gives the real parameter $q$ a time-dependent phase:
\begin{equation}
  \label{eq:249}
  q_0=e^{\ds -\epsilon/2} \to q(t)=e^{\ds-\epsilon(1+i\omega t)/2}=q_0\,e^{\ds-i\epsilon\,\omega
 t/2}\,,~\text{ i.e.\ } \tau_0=\frac{\epsilon\,i}{2\pi} \to \tau(t)=\tau_0\,(1+i\,\omega t)\,.
\end{equation}
Notice that for $\dl=0$ only the parameter $q$ remains time-dependent!

The above expressions show that a non-vanishing $\dl$ leads to  non-trivial complications
of the time evolution for the states \eqref{eq:101} and \eqref{eq:162}! 

 The propagator \eqref{eq:309} was first analysed by Schulman \cite{schul1} and
for the special cases $\epsilon =1$,\,$\dl=0 \text{ and } \dl=0.5$ the time evolution of the
states \eqref{eq:162} was discussed  by Kowalski and Rembieli\'{n}ski \cite{kow3}.
\section*{Acknowledgements}
Part of this work was done during a 2-month stay in the last summer
 at the Albert-Einstein Institute for Gravitation in Potsdam. I am very grateful to its
Director Hermann Nicolai  and to
Thomas Thiemann for the kind invitation to come there and I thank Thomas Thiemann and Martin
Bojowald for stimulating discussions. I also thank the Theory Group of DESY, Hamburg, for
its enduring very helpful and generous hospitality after my retirement from the Institute for
Theoretical Physics of the RWTH Aachen. Last but not least I am deeply obliged to my wife
Dorothea for her enormous patience and important support!

I thank K.\ Kowalski for remarks on the first version of this paper and D.\ Trifonov for
comments on a later one. Finally I thank the anonymous referee for suggesting
 $T$-violation by means
of an external OAM $L_{\vp}^{ext}$ [see the third paragraph after Eq.\ \eqref{eq:303}].

 \newpage
\section*{Appendices}
\appendix \section{The ``fault'' of the angle}
The present appendix summarizes the arguments, why the angle $\vp$ itself it not a good
global observable on the phase space \eqref{eq:12}, neither classically nor quantum theoretically
and why it should be replaced by the functions $\cos\vp$ and $\sin\vp$!

 Let me start with a
well-known classical example, which illustrates the main point:

Consider an infinitely thin arbitrarily long straight conducting wire extending along
 the $z$-axis of
a rectangular coordinate system and having the charge density $\sigma$ per unit of length.
In the punctured $(x,y)$-plane orthogonal to the wire we have the electric field
\begin{equation}
  \label{eq:250}
  \vec{E}(\vec{x}) =\frac{\sigma}{2\pi\epsilon_0}\,\frac{\vec{x}}{x^2+y^2}\,,~\vec{x}=(x,y)\,,
~~\mbox{div}\vec{E}=0\,,~
\mbox{curl}\vec{E}=0\,,\mbox{ on } \mathbb{R}^2-\{0\}\,.
\end{equation}
The field \eqref{eq:250} may be derived from a potential $\phi_e$:
\begin{equation}
  \label{eq:251}
  \vec{E}=\, \mbox{grad}\phi_{e}(r)\,,~~\phi_{e}(r)=\frac{\sigma}{2\pi\epsilon_0}\,
\ln\left(\frac{r}{r_0}\right)\,,
~~r=\sqrt{x^2+y^2}\,,\,r_0 \in \mathbb{R}^2-\{0\}\,,
\end{equation}
where $\phi_{e}(r)$ is a well-defined smooth function on the punctured plane
 $\mathbb{R}^2-\{0\}$
with the following global property:

If $C_{1\to 2}$ is any smooth path in the punctured $(x,y)$-plane from a point $P_1$ to a
point $P_2$, then the potential difference
\begin{equation}
  \label{eq:252}
  \phi_{e}(P_2)-\phi_{e}(P_1)=\int_{C_{1\to 2}}d\phi_{e}=\int_{C_{1\to 2}}
\partial_x\phi_{e}(\vec{x})\,dx + \partial_y\phi_{e}(\vec{x})\,dy
\end{equation}
is uniquely defined even if the path $C_{1\to2}$ circles the charged wire several times!

Next, consider another infinitely thin arbitrarily long straight wire along the $z$-axis
through which a constant electric current $I$ flows in the positive $z$-direction. The current
generates a magnetic field in the punctured $(x,y)$-plane of the form
\begin{equation}
  \label{eq:253}
  \vec{B}(\vec{x}) = \frac{\mu_0\,I}{2\pi}\,\frac{1}{x^2+y^2}(-y,x)\,,~~
\mbox{div}\vec{B}=0\,,~
\mbox{curl}\vec{B}=0\,,\mbox{ on } \mathbb{R}^2-\{0\}\,.
\end{equation}
If we introduce polar coordinates
\begin{equation}
  \label{eq:254}
  x=r\,\cos\vp\,,~~y=r\,\sin\vp\,,
\end{equation}
then we have
\begin{equation}
  \label{eq:255}
  \vec{B}(\vec{x})\cdot(dx,dy) = \frac{\mu_0\,I}{2\pi}\,d\vp\,\,\mbox{ on }\mathbb{R}^2-\{0\}\,.
\end{equation}
This suggests to introduce a scalar magnetic potential \cite{pan} by
\begin{equation}
  \label{eq:256}
\vec{B}= \mbox{grad}\phi_{m}\,,~~\phi_m(\vec{x})= \frac{\mu_0\,I}{2\pi}\,\arctan\left(\frac{y}{x}
\right)\,,~~d\phi_m = \frac{\mu_0\,I}{2\pi}\,d\vp\,,~\vp \in \mathbb{R} \bmod{2\pi}\,.
\end{equation}
Here, however, we encounter a problem:
It follows from 
\begin{equation}
  \label{eq:257}
  \mbox{curl}\vec{B}(\vec{x})=\mu_0\,I\,\dl(\vec{x})\,\vec{e}_z
\end{equation}
that
\begin{equation}
  \label{eq:258}
 \oint_{r=a}\vec{B}(\vec{x})(dx,dy) =2\pi\,a\,B_{\vp}=\mu_0\,I\,,~~B_{\vp} =\frac{\mu_0\,I}{
2\pi\,a}\,, 
\end{equation}
where $B_{\vp}\,(-\sin\vp,\cos\vp)$ is the unique magnetic field tangential to the
 circle of radius $a$ at $\vec{x}=
a(\cos\vp,\sin\vp)$\,. If we now again consider a smooth path $C_{1\to2}$ from a point $P_1$ to
a point $P_2$, both in $\mathbb{R}^2-\{0\}$, then the value of the  integral
\begin{equation}
  \label{eq:259}
  \int_{C_{1\to2}} \vec{B}(\vec{x})\cdot(dx,dy)=\int_{C_{1\to2}}
d\phi_m = \frac{\mu_0\,I}{2\pi}\,\int_{C_{1\to2}}d\vp
\end{equation}
is no longer uniquely defined!
The integral over the angle $\vp$ gives  the correct physical magnetic field $B_{\vp}$ only
if we restrict $\vp$ to the interval $[0,2\pi)$! If the path $C_{1\to2}$ circles the current
twice, we would get for $B_{\vp}$ twice its physical value and so on. The point is that 
- contrary to $\phi_e$ from above - the ``potential'' $\phi_m(\vec{x})$ is not a globally
 well-defined function on  $\mathbb{R}^2-\{0\}$, because the angle $\vp$ is not one: When
$\vp$ reaches the value $2\pi$ it has to ``jump back'' to $0$, i.e.\ it has a discontinuity.
In textbooks (see e.g.\ Ref.\ \cite{pan}) for electrodynamics this behaviour is compared to the
 above electrostatic case
with an additional infinitely thin electric dipole sheet which causes a corresponding
discontinuity for $\phi_e$ if one passes the sheet. This, however, is a physical effect whereas
the discontinuity of $\phi_m$ is due to a complication as to its mathematical properties:

Mathematically speaking, the (exterior) differential 1-form
\begin{equation}
  \label{eq:260}
  \gamma(\vec{x}) =\frac{1}{x^2+y^2}(-y\,dx +x\,dy)=``(d\vp)''\mbox{ on }\mathbb{R}^2-\{0\}
\end{equation}
is a {\em closed} form but {\em not} an {\em exact} one, i.e.\ we have
\begin{equation}
  \label{eq:261}
  d\gamma =0\,,
\end{equation}
but $\gamma$ {\em cannot} be represented as $\gamma(\vec{x}) =df(\vec{x})$, where $f(\vec{x})$
is a smooth function {\em globally} well-defined on $\mathbb{R}^2-\{0\}$! 

The 1-form \eqref{eq:260} is the standard example in textbooks (see, e.g.\ \cite{gui,thirr})
for a closed differential form which is not an exact one! The difference signals that the
manifold on which the closed differential form is defined has a non-trivial global topological
structure. In our case it is the punctured plane $\mathbb{R}^2-\{0\}$ which is not simply
 connected and therefore {\em globally non-trivial}.

In section 2 I stressed under number v) (around Eq.\ \eqref{eq:54}) that for the group 
theoretical quantization procedure to succeed one needs {\em globally} well-defined Hamiltonian
functions on the phase space. This is not the case for the angle $\vp$ (see also Appendix B),
but it is so for the two periodical functions $\cos\vp$ and $\sin\vp$ which are smooth and the
knowledge of which allows to determine the associated $\vp \in [0,2\pi)$ uniquely!

The difficulties with the angle $\vp$ on the classical level persist in the quantum theory:
As to the details of the following mathematical sketches see the excellent discussions
by Robinson \cite{rob1} and Reed and Simon \cite{reed1}!
 
On any open interval $(\vp_1,\vp_2) \subset [0,2\pi]$ we have
\begin{equation}
  \label{eq:262}
  [\vp,\tilde{L}] = i\,\, \mbox{ for } \vp \in (\vp_1,\vp_2) \subset [0,2\pi]\,,~\vp
 \in \mathbb{R} \bmod{2\pi}\,,~~\tilde{L}=\frac{1}{i}\partial_{\vp}\,.
\end{equation}
Here $\vp$ appears as a multiplication operator on $L^2([0,2\pi],d\vp/2\pi)$ with the scalar
product \eqref{eq:22}. The differential operator $\tilde{L}$
in general may act on absolutely continuous functions $f(\vp)$ on that space, i.e.\ on functions
which allow for a representation
\begin{equation}
  \label{eq:263}
  f(\vp) = \int_0^{\vp}d\phi\,g(\phi) + \mbox{ const. }\,.
\end{equation}
If one - formally - assumes the commutator \eqref{eq:262} to hold in general, one 
immediately encounters a
contradiction:
\begin{equation}
  \label{eq:264}
  (e_m,[\vp,\tilde{L}]\,e_n)=(e_m,\vp\,\tilde{L}\,e_n)-(\tilde{L}e_m,\vp\, e_n) =
(n-m)\,(e_m,\vp\,e_n)=i\,\dl_{mn}\,, 
\end{equation}
which gives $0=i$ for $m=n$!.

The background of this difficulty is that $\vp$ is not  differentiable at the boundaries
of $[0,2\pi]$, so that $\tilde{L}$ is not applicable to $\vp f(\vp)$ there. One might try
to avoid this difficulty by restricting oneself to functions $h(\vp)$ with the boundary
 properties
 \begin{equation}
   \label{eq:265}
   h(0)=0=h(2\pi)\,.
 \end{equation}
But now $\tilde{L}$ is merely symmetric (i.e.\ $(h_2,\tilde{L}\,h_1)=(\tilde{L}\,h_2,h_1)$)
 on this set (domain) of functions, not self-adjoint,
 i.e.\ it has no satisfactory 
spectral decomposition (see the Refs.\ mentioned above). This can already be inferred from
the fact that the functions $e_n(\vp)$ from Eq.\ \eqref{eq:28} do not obey the boundary condition
\eqref{eq:265}! But this symmetric $\tilde{L}$ has a 1-parametric set of self-adjoint extensions
to the space of functions $\psi(\vp)$ which obey the boundary condition \eqref{eq:42}! The
 parameter $\dl$ there also characterizes the self-adjoint extensions $\tilde{L}_{\dl}$ of
 $\tilde{L}$ we encountered in Sec.\ 1 in a different context. 

There have been many attempts to find cures for the difficulties indicated by the relation
\eqref{eq:264}:

One is to allow - reluctantly - for $\dl$-functions at the boundaries 
\cite{jud,susk,carr,barn3,barn4}. This in general will
destroy self-adjointness in the usual understanding of Hilbert space operators. \\ Another is
to use only finite-dimensional vector spaces of dimension $d$, calculate the physical quantities
like expectation values etc.\ and let $d$ go to infinity at the very end \cite{barn}. This
 procedure has its problems, too: E.g.\ assume that the operators $\vp$ and $\tilde{L}$ with
the commutator \eqref{eq:262} may be represented by finite-dimensional matrices in a
 $d$-dimensional vector space.  Then taking the trace of both sides of the commutator relation 
\eqref{eq:262} yields the contradiction $0=i\,d$ (because $\tr{A\,B})=\tr{B\,A}$)). So one has to
take care of this new problem by modifying the commutator. In addition,
 in finite-dimensional vector spaces there is no
difference between symmetrical (hermitean) and self-adjoint operators, a difference which is
important in infinite-dimensional Hilbert spaces, because one would like to have a decent
spectral decomposition. Thus, there appear to be problems with the limit $d\to \infty$ 
\cite{dub,fuj}. Finally, there seems to be no chance to derive the existence of quasi-OAM
$\dl$ by starting from finite dimensional vector spaces!

Other authors \cite{chi,forb,trif} discussed related problems associated with uncertainty
 relations for wave functions on the circle.

All these problems can be avoided by using the functions $\cos\vp$ and $\sin\vp$ as basic
 observables - instead of $\vp$ itself - with their
algebraic structure \eqref{eq:16} which constitutes the Lie algebra of the euclidean group
$E(2)$ and all its covering groups!
\section{On the group theoretical quantization of the phase space
 $S^1\times \mathbb{R}$}
Basic ingredients for quantizing the phase space \eqref{eq:12} in terms of irreducible
unitary representations of the euclidean group $E(2)$ have already been discussed in the
Introduction and in Sec.\ 2. As to general introductions to the concept of group theoretical
quantization see the excellent article by Isham \cite{ish} and the similarly excellent book
by Guillemin and Sternberg \cite{stern}.

 The euclidean group $E(2)$ and its covering groups
with their irreducible unitary representations plays a prominent role in Isham's  discussions
 as an example for the new quantum effects induced by non-trivial topologies like that of
the configuration space $S^1$, namely the existence of non-vanishing $\dl$-effects!
 The Lie algebra  $\mathfrak{e}(2)$ of the euclidean group $E(2)$ is also discussed by Guillemin
and Sternberg in a different context \cite{ster2}. An appealing introduction into the irreducible
representations of the group $E(2)$ itself can be found in Ch.\ IV of Sugiura's book \cite{Su},
including Plancherel's theorem (``Fourier'' transform) for that group. Sugiura also gives a
nice introduction to the concept of ``induced representations'' which provides the appropriate 
method to construct the irreducible unitary representations of $E(2)$. 

The analogue of the Groenewold - Van Hove obstruction for the conventional quantization procedure
is discussed in case of the present phase space  $S^1 \times \mathbb{R}$ and its quantization
 in terms of the canonical group $E(2)$  in Ref.\ \cite{got}.

Let me briefly recall the main structure of the group $E(2)$: It is convenient to use complex
coordinates $z=x+iy$ on the plane $\mathbb{R}^2$. The euclidean scalar product of two vectors
$z_1$ and $z_2$ may be written as
\begin{equation}
  \label{eq:266}
  (z_2,z_1)\equiv x_2\,x_1+y_2\,y_1 = \Re{(z_2^*\cdot z_1)}\,.
\end{equation}
The euclidean group $E(2)$ consists of all linear transformations of the plane which leave
the square of the distance
\begin{equation}
  \label{eq:267}
  (z_2-z_1,z_2-z_1)=(x_2-x_1)^2 +(y_2-y_1)^2
\end{equation}
invariant. If we confine ourselves to those transformations which are continuously connected
to the identity transformation (i.e.\ we exclude reflections $z \to z^*$), we have
\begin{eqnarray}
  \label{eq:268}
  \text{ translations } T_2(t)\,&:&~~z\to z+t\,,~t=a+ib\,,~a,b \in \mathbb{R}\,,\\
\text{ rotations } R(\alpha)\,&:&~~z\to e^{\ds i\alpha}\,z\,,~\alpha \in \mathbb{R} \bmod{2\pi}
\,. \label{eq:269}
\end{eqnarray}
If we define
\begin{equation}
  \label{eq:270}
  g(\alpha,t)=\begin{pmatrix}e^{\ds i\alpha}&t \\0&1 \end{pmatrix}\,,
\end{equation}
we can combine the two transformations \eqref{eq:268} and \eqref{eq:269} into
\begin{equation}
  \label{eq:271}
  g(\alpha,t)\,\begin{pmatrix}z\\1 \end{pmatrix}=\begin{pmatrix}e^{\ds i\alpha}z+t \\1
 \end{pmatrix}\,.
\end{equation}
 The group multiplication law is
\begin{equation}
  \label{eq:272}
  g(\alpha_2,t_2)\circ g(\alpha_1,t_1)=g(\alpha_1+\alpha_2\,\bmod{2\pi},t_2+e^{\ds i\alpha_2}
\,t_1)=g(\alpha_3 \bmod{2\pi},t_3)\,.
\end{equation}
When applying the group element $g(\alpha,t)$ to a point $s=(\vp,p_{\vp})$ of the phase
 space \eqref{eq:12} we write the group parameters $\alpha$ and $t$ as indices. According to
Eq.\ \eqref{eq:55} the action of the group $E(2)$ on the phase space \eqref{eq:12} is given by
\cite{ish2}
\begin{equation}
  \label{eq:281}
  g_{\alpha,t}(s) =(\vp',p_{\vp}')=[(\vp+\alpha)\bmod{2\pi},\, p_{\vp}+ a\, \sin
(\vp+\alpha) -b\,\cos(\vp +\alpha)]\,.
\end{equation}
The transformation is obviously symplectic, i.e.\ we have
\begin{equation}
  \label{eq:282}
  d\vp'\wedge dp_{\vp}'=d\vp \wedge dp_{\vp}\,.
\end{equation}
It is also transitive, i.e.\ given any two points $s_1$ and $s_2$ there is always
 a transformation \eqref{eq:281} which transforms $s_1$ into $s_2$: The choice $\alpha=\vp_2-
\vp_1$ transforms $\vp_1$ into $\vp_2$. The remaining requirement
\begin{equation}
  \label{eq:283}
  p_{\vp,2}-p_{\vp,1}=a\sin\vp_2-b\cos\vp_2
\end{equation}
can be fulfilled by an appropriate choice of $a$ and $b$.

The action is almost effective, i.e.\ it follows from
\begin{equation}
  \label{eq:284}
  g_{\alpha,t}(s) = s\, \forall\, s\,,
\end{equation}
that 
\begin{equation}
  \label{eq:285}
  \alpha = 2\pi n\,,~n \in \mathbb{Z}\,,~t=0\,.
\end{equation}
This represents the center $\mathbb{Z}$ of the universal covering group $\widetilde{SO(2)}
=\mathbb{R}$ of the
rotation group  $SO(2)$.
A special solution of the condition \eqref{eq:284} is
\begin{equation}
  \label{eq:286}
  \alpha = \alpha \in \mathbb{R} \bmod{2\pi q}\,,~q \in \mathbb{N}\,,
\end{equation}
which represents the center $Z_q$ of  the $q$-fold covering group of $SO(2)$. It may be 
explicitly implemented by replacing the angle $\alpha$ in \eqref{eq:272} by $\beta =\alpha/q\,,
\beta \in \mathbb{R} \bmod{2\pi}$. The group law for the universal covering law can be given by
the relation \eqref{eq:272} by omitting the condition $\bmod{2\pi}$ completely.

The other conditions for the group $E(2)$ to be the canonical (quantizing) group for the
phase space \eqref{eq:12} have been discussed in Sec.\ 2.

Let us denote by $T_1(a)$ and $T_1(b)$ the one-dimensional translation subgroups in $x$- and
$y$-direction, respectively. Then it follows from \eqref{eq:281} that the subgroup $T_1(a)$
leave the points of the two lines $\vp=0$ and $\vp=\pi$ unchanged (``stable'')
 (if $\alpha =0,\,b=0$) and
that the subgroup $T_1(b)$ does the same with the lines $\vp=\pi/2$ and $\vp =3\pi/2$. Thus,
we may describe the phase space \eqref{eq:12} as one of the homogeneous spaces
\begin{equation}
  \label{eq:288}
  \mathcal{S}_{\vp,p_{\vp}} \cong E(2)[\alpha,t]/T_1(a)\cong E(2)[\alpha,t]/T_1(b)\,.
\end{equation}

The euclidean group $E(2)$ also plays a role in several papers on Moyal $\star$-product
 (``deformation'')
quantization of the phase space \eqref{eq:12} \cite{frons1,frons2,alc1,alc2,ple,gon3}.

The occurence of inequivalent irreducibble unitary representations \eqref{eq:24} characterized
by the parameter $\dl$ has its correspondence in inequivalent representations of the Weyl-algebra
generated by $U=\exp{i\alpha\vp}$ and $V=\exp{i\beta \tilde{L}}$ \cite{morch}.

\section{Some properties of $\vt$-functions}
In Secs.\ 4 and 5 above on the coherent states of the circle Jacobi's $\vt$-functions play
 a prominent
role. I shall briefly mention some suitable textbooks where one can find their appropriate
 properties and shall add a few relations here.
One inconvience as to the literature is that different authors use different conventions
for the arguments of the $\vt$-functions. Take
\begin{equation}
  \label{eq:287}
  \vt_3^{(a)}(\zeta,q=e^{\ds i\pi\tau}) =\sum_{n \in \mathbb{Z}}q^{\ds n^2}e^{\ds 2ina\zeta}\,.
\end{equation}
Some authors have $a=1$, others $a=\pi$. I have used the convention $a=1$ of the very
 useful book
by Whittaker and Watson \cite{whit}. The same convention have the ``classical'' introductory 
book by
Bellman \cite{bell} and the more recent appealing textbook by Lawden \cite{lawd}. 
Erd\'{e}lyi et al.\ have $a=\pi$, so has Mumford's influential modern textbook \cite{mum1}. 
Useful are also the formulae in Ref,\ \cite{abr1} which has $a=1$.
Most of the  formulae concerning the $\vt$-functions needed in the text above have been given
there. As the ratios of $\vt$-functions are related to Jacobi's elliptic functions, one may
express their ratios in Eqs.\  \eqref{eq:172}, \eqref{eq:184}, \eqref{eq:193} and 
\eqref{eq:195} in
terms of elliptic functions $\text{sn}(u,k),\,\text{cn}(u,k)$ and $\text{dn}(u,k)$ (for their 
definition see Refs.\ \cite{whit,lawd}):
\begin{equation}
  \label{eq:293}
  \frac{\vt_4(\zeta,q)}{\vt_3(\zeta,q)}=\frac{\sqrt{k'}}{\dn(u,k)}\,,~u=\frac{2K}{\pi}\zeta\,,~
k=\frac{\vt_2^2(\zeta=0,q)}{\vt_3^2(\zeta=0,q)}\,,~
k'=\frac{\vt_4^2(\zeta=0,q)}{\vt_3^2(\zeta=0,q)}\,,~\frac{2K}{\pi}=\vt_3^2(\zeta=0,q)\,.
\end{equation}
Furthermore
\begin{equation}
  \label{eq:294}
  \frac{\vt_3^{\prime}(\zeta,q)}{\vt_3(\zeta,q)}= \frac{\vt_4^{\prime}(\zeta,q)}{\vt_4(\zeta,q)}
-\frac{2K}{\pi}\,k^2\,\frac{\cn(u,k)\,\sn(u,k)}{\dn(u,k)}\,,
\end{equation}
where 
\begin{eqnarray}
  \label{eq:295}
  \frac{\vt_4^{\prime}(\zeta,q)}{\vt_4(\zeta,q)}&=&\frac{2K(k)}{\pi}\,Z(u)\,,~Z(u)=
[E(u,k)-u\,E(k)/K(k)]\,, \\
E(u,k)&=& \int_0^u dv\,\dn^2(v,k)\,, \nonumber \\
E(k)&=&E(u=K,k)=\int_0^{\pi/2}d\phi\, (1-k^2\sin^2\phi)^{1/2}\,,\nonumber \\ 
 K(k)&=&\int_0^{\pi/2}d\phi\,
 (1-k^2\sin^2\phi)^{-1/2}\,.\nonumber
\end{eqnarray}
$Z(u)$ is Jacobi's ``Zeta-Funktion''.
Finally
\begin{equation}
  \label{eq:296}
  \frac{\vt_3^{\prime \prime}}{\vt_3}-\frac{\vt_3^{\prime\,2}}{\vt_3^2}=\frac{d^2\ln\vt_3(
\zeta,q)}{d\zeta^2}=\frac{4K^2(k)}{\pi^2}\left[\frac{k^{'2}}{\dn^2(u,k)}-E(k)/K(k)\right]\,,~
k^{'2}+k^2=1\,.
\end{equation}

\end{document}